\pdfoutput=1
\documentclass[11pt,a4paper]{article}
\usepackage[utf8]{inputenc}
\usepackage{jheppub}
\hypersetup{unicode=true,bookmarksopen=true}

\usepackage{amsmath}
\usepackage{mathtools}
\usepackage{lmodern}
\usepackage[T1]{fontenc}
\usepackage{bbm}
\DeclareMathAlphabet{\mathbfi}{OML}{cmm}{b}{it}

\usepackage{feynmp-auto}
\usepackage{subcaption}

\let\originalleft\left
\let\originalright\right
\renewcommand{\left}{\mathopen{}\mathclose\bgroup\originalleft}
\renewcommand{\right}{\aftergroup\egroup\originalright}

\makeatletter
\newenvironment{equations}[1][]{\subequations\ifx\relax#1\relax\else\label{#1}\fi\align\ignorespaces}{\endalign\ignorespacesafterend\endsubequations}
\def\@spliteq#1{\begin{equation}\begin{split}#1\end{split}\end{equation}}
\def\splitequation{\collect@body\@spliteq}

\makeatother

\renewcommand{\vec}[1]{{\ifnum9<1#1\mathbf{#1}\else\ifcat\noexpand#1\relax\boldsymbol{#1}\else\mathbfi{#1}\fi\fi}}
\newcommand{\mathe}{\mathrm{e}}
\newcommand{\mathi}{\mathrm{i}}
\let\oldre\Re
\let\oldim\Im
\renewcommand{\Re}{\oldre\mathfrak{e}\,}
\renewcommand{\Im}{\oldim\mathfrak{m}\,}
\newcommand{\total}{\mathop{}\!\mathrm{d}}

\newcommand{\abs}[1]{{\left\lvert{#1}\right\rvert}}

\newcommand{\tr}{\operatorname{tr}}

\newcommand{\eqend}[1]{\,#1}
\newcommand{\bigo}[1]{\mathcal{O}\left({#1}\right)}
\newcommand{\brst}{\mathop{\mathsf{s}}}

\makeatletter
\def\expect{\@ifnextchar[{\@expecttw@}{\@expect@ne}}
\def\@expecttw@[#1]#2{\left\langle{#2}\right\rangle^\text{#1}}
\def\@expect@ne#1{\left\langle{#1}\right\rangle}
\makeatother

{%
\end{oldthebibliography}%
}

\allowdisplaybreaks

\title{Graviton corrections to the Newtonian potential using invariant observables}

\author{M.~B.~Fröb,}
\author{C.~Rein}
\author{and R.~Verch}

\affiliation{Institut f{\"u}r Theoretische Physik, Universit{\"a}t Leipzig,\\ Br{\"u}derstra{\ss}e 16, 04103 Leipzig, Germany}

\emailAdd{mfroeb@itp.uni-leipzig.de}
\emailAdd{rein@itp.uni-leipzig.de}
\emailAdd{rainer.verch@uni-leipzig.de}

\abstract{We consider the effective theory of perturbative quantum gravity coupled to a point particle, quantizing fluctuations of both the gravitational field and the particle's position around flat space. Using a recent relational approach to construct gauge-invariant observables, we compute one-loop graviton corrections to the invariant metric perturbation, whose time-time component gives the Newtonian gravitational potential. The resulting quantum correction consists of two parts: the first stems from graviton loops and agrees with the correction derived by other methods, while the second one is sourced by the quantum fluctuations of the particle's position and energy-momentum, and may be viewed as an analog of a ``Zitterbewegung''. As a check on the computation, we also recover classical corrections which agree with the perturbative expansion of the Schwarzschild metric.}

\begin{document}

\maketitle

\section{Introduction}
\label{sec_introduction}

Irrespective of an eventual UV completion, one can describe quantum gravitational phenomena well below some cutoff scale (usually taken to be the Planck scale) using an effective quantum field theory of gravity~\cite{donoghue1994b,burgess2004}. In this approach, one perturbatively quantizes metric fluctuations around a fixed background, with the action given by an expansion of the usual Einstein--Hilbert action, possibly coupled to matter fields. One important prediction that can be made in this way are quantum corrections to the Newtonian gravitational potential, which have been studied by many authors~\cite{radkowski1970,schwinger1968,duff1974,capperduffhalpern1974,capperduff1974,donoghue1994a,donoghue1994b,muzinichvokos1995,hamberliu1995,akhundovbelluccishiekh1997,dalvitmazzitelli1997,duffliu2000a,duffliu2000b,kirilinkhriplovich2002,khriplovichkirilin2003,bjerrumbohrdonoghueholstein2003a,bjerrumbohrdonoghueholstein2003b,satzmazzitellialvarez2005,holsteinross2008,parkwoodard2010,marunovicprokopec2011,marunovicprokopec2012,burnspilaftsis2015,bjerrumbohretal2016,froeb2016}. The potential including one-loop quantum corrections has the form
\begin{equation}
\label{intro_newtonpotential}
V(r) = - \frac{G_\text{N} m}{r} \left( 1 + c \frac{\ell_{\text{Pl}}^2}{r^2} \right) \eqend{,}
\end{equation}
where $m$ is the mass of the source, $c$ is a numerical constant that depends on the type of massless fields\footnote{For massive fields, there are additional contributions that fall-off exponentially with $r$ and one obtains a Yukawa-type potential.} that run in the loop, and $\ell_{\text{Pl}} = \sqrt{ \hbar G_\text{N}/c^3 }$ the Planck length. By dimensional considerations, it is clear that the result must take this form, and is thus experimentally insignificant. For corrections induced by matter fields, the result for $c$ is unambiguous, while for gravitons it depends on the method that is used to compute the corrections and define the potential. This stems from the fact that perturbative quantum gravity is a gauge theory, with infinitesimal diffeomorphisms as gauge transformations, but that in contrast to gauge theories of Yang--Mills type the gauge symmetry is external, i.e., a gauge transformation of a quantity $A(x)$ is given by the Lie derivative $\mathcal{L}_\xi A(x)$. It follows that any field defined at a fixed point $x$ of the background spacetime cannot be invariant, and any invariant observable must necessarily be non-local, in contrast to Yang--Mills theories where for example $\tr F^2(x)$ is a perfectly sensible invariant observable. If one determines the Newtonian potential from an inverse scattering analysis, the value for $c$ is given by
\begin{equation}
\label{intro_constant_c}
c = \frac{41}{10 \pi} + \frac{[ 1 + \frac{5}{4} (1-6\xi)^2 ] N_0 + 6 N_{1/2} + 12 N_1}{45 \pi} \eqend{,}
\end{equation}
where the first term is the contribution from gravitons, and the second term the contribution from $N_0$ scalars with non-minimal curvature coupling $\xi$, $N_{1/2}$ spinors and $N_1$ gauge fields~\cite{kirilinkhriplovich2002,khriplovichkirilin2003,bjerrumbohrdonoghueholstein2003a,bjerrumbohrdonoghueholstein2003b,bjerrumbohrdonoghuevanhove2014}.

However, the result~\eqref{intro_newtonpotential} can change qualitatively once one considers non-trivial backgrounds. While for a generic background spacetime no scattering matrix exists, and the inverse scattering method is not available, one can compute the same corrections by solving the effective field equations for the metric perturbations obtained from varying an effective action where matter fluctuations have been integrated out~\cite{duff1974,duffliu2000a,duffliu2000b,parkwoodard2010,marunovicprokopec2011,marunovicprokopec2012,froeb2016}. In flat space, this approach gives the same results~\eqref{intro_constant_c} for $c$ (excluding gravitons), and it has also been generalized to de~Sitter space~\cite{wangwoodard2015b,parkprokopecwoodard2016,froebverdaguer2016a,froebverdaguer2017}. Since de~Sitter spacetime is not only a very good approximation to the inflationary period in the standard cosmological model~\cite{mukhanov,kazanas1980,sato1981,guth1981,linde1982,albrechtsteinhardt1982}, but also models our present accelerated universe~\cite{perlmutteretal1998,knopetal2003,eisensteinetal2005,astieretal2006,kowalskietal2008}, obtaining quantum corrections in this background is important. Besides the Planck length $\ell_{\text{Pl}}$, the inverse Hubble rate $H^{-1}$ furnishes another length scale, such that from dimensional analysis alone different corrections become possible. The explicit computations show that additional contributions arise which grow logarithmically with either time or distance, and are thus potentially much larger than in flat space (even though they are still out of reach of experimental verification).

It is thus important to try and incorporate also effects from graviton loops into the latter method, and in particular to verify whether those can grow observationally large. At linear order in metric perturbations one can use the Bardeen variables~\cite{bardeen1980} to obtain a gauge-invariant Newtonian potential, and the inclusion of matter loops into the effective action is straightforward. When one wants to obtain the effects of graviton loops, however, one leaves the realm of linear perturbations and has to include non-linear self-interactions of the graviton, whence the Bardeen variables are not invariant observables anymore. To determine a gauge-invariant observable corresponding to the Newtonian potential, we follow instead the recent proposal~\cite{brunettietal2016} and its extensions~\cite{froeb2018,froeblima2018,froeb2019,lima2020}, where relational observables for the gravitational field were constructed to all orders in perturbation theory, and which at linear order reduce to the Bardeen variables. As with all relational approaches, these observables are given by the value of one dynamical field of the theory at the point where another dynamical field has a given value, i.e., the value of one field in relation to a second one. While relational observables have a long history~\cite{komar1958,bergmannkomar1960,bergmann1961,gieseletal2010a,gieseletal2010b,tambornino2012}, the important breakthrough of~\cite{brunettietal2016} was the construction of such observables for highly symmetric backgrounds without the introduction of extra fields (such as the Brown--Kucha{\v r} dust~\cite{brownkuchar1995}) which change the physics.

Summarising, the aim of the calculation that we present here is the computation of graviton loop corrections to the Newtonian potential by a new method: instead of using the inverse scattering method (i.e., by computing the one-loop $S$-matrix element corresponding to the scattering of scalar fields, and determining the non-relativistic potential that gives the same $S$-matrix element in the Born approximation), we compute the expectation value of an invariant observable corresponding to the metric perturbation, sourced by a point particle. For matter loop corrections, it is known that the results of both methods agree, and we want to check whether the same holds for graviton loop corrections. The main advantage of the latter method is that it may be generalised to curved backgrounds, where no $S$-matrix can be defined.

Our calculation is organized as follows: first we construct the gauge-invariant observable corresponding to the Newtonian potential in section~\ref{sec_invobs}, then expand the action for gravity coupled to a point particle to the order required for a one-loop computation in section~\ref{sec_action}, and deal with the gauge invariance (both infinitesimal diffeomorphisms and wordline reparametrizations) using the BRST method in section~\ref{sec_brst}. We then compute the expectation value of this observable to one-loop order in section~\ref{sec_correction}, and separate the various contributions into corrections coming from graviton and ghost loops in section~\ref{sec_loop}, corrections coming from the fluctuation of the particle's worldline in section~\ref{sec_worldline}, higher-order classical corrections in section~\ref{sec_class} and corrections due to the relational observable in section~\ref{sec_coord}. Since the result is gauge-invariant by construction, a judicious choice of gauge fixing vastly simplifies the actual computations, and we explain which corrections vanish by this choice in section~\ref{sec_vanish}. We use the ``+++'' convention of Ref.~\cite{mtw}, work in $n$ dimensions to use dimensional regularization, choose $c = \hbar = 1$, and define $\kappa^2 \equiv 16 \pi G_\text{N}$ with Newton's constant $G_\text{N}$. Greek indices range over spacetime, while Latin ones are purely spatial.

\section{Observables, actions and BRST invariance}
\label{sec_obsactionbrst}

We take Minkowski spacetime with Cartesian coordinates $x^\mu$ and flat metric $\eta_{\mu\nu}$ as our background spacetime. We then consider metric perturbations $h_{\mu\nu}$ in this spacetime, such that the full metric reads
\begin{equation}
\label{eq_metric_expansion}
g_{\mu\nu} = \eta_{\mu\nu} + \kappa h_{\mu\nu} \eqend{.}
\end{equation}
An infinitesimal diffeomorphism $\delta_\xi x^\mu = - \kappa \xi^\mu(x)$ changes the metric according to $\delta_\xi g_{\mu\nu} = \kappa \mathcal{L}_\xi g_{\mu\nu}$, where $\mathcal{L}$ is the Lie derivative, and accordingly the metric perturbation changes as
\begin{equation}
\label{eq_metric_transform}
\delta_\xi h_{\mu\nu} = \mathcal{L}_\xi \eta_{\mu\nu} + \kappa \mathcal{L}_\xi h_{\mu\nu} = 2 \partial_{(\mu} \xi_{\nu)} + \kappa \xi^\rho \partial_\rho h_{\mu\nu} + 2 \kappa h_{\rho(\mu} \partial_{\nu)} \xi^\rho \eqend{.}
\end{equation}
Similarly, the change of any tensor in the perturbed geometry is given by its Lie derivative with respect to $\xi$. The Stewart--Walker lemma~\cite{stewartwalker1974} shows that the only local gauge-invariant observables that one can construct at linear order in perturbations are obtained from tensors whose background value is a constant linear combination of $\delta_\mu^\nu$'s.\footnote{While the Bardeen variables are gauge-invariant at linear order, they are not local.} For a Minkowski background spacetime, the linearized Riemann tensor is such a local invariant observable, and in fact a complete set of linearized gauge-invariant observables can be obtained from a so-called IDEAL classification of spacetimes~\cite{canepadappiaggikhavkine2018,froebhackkhavkine2018,khavkine2019}. However, these observables become gauge-dependent at second and higher orders, which has a clear physical reason: since diffeomorphisms move points, no local observable defined at a single point can be invariant, and all invariant observables must be necessarily non-local. In the next subsection, we will construct such non-local observables.

\subsection{Gauge-invariant observables in perturbative gravity}
\label{sec_invobs}

Following Refs.~\cite{brunettietal2016,froeb2018,froeblima2018,froeb2019,lima2020}, we construct gauge-invariant observables in a relational way. First, one needs to define field-dependent coordinates that transform as scalars under diffeomorphisms of the background spacetime coordinates. The gauge-invariant observables are then obtained by a field-dependent diffeomorphism from the background spacetime coordinates to the field-dependent ones, evaluating the quantity of interest in the new coordinate system. There are many ways to obtain the necessary field-dependent coordinates, with the earliest proposals~\cite{komar1958,bergmannkomar1960,bergmann1961} being various curvature scalars. However, for a highly symmetric background such as Minkowski space, these all vanish and do not actually discriminate points. The important breakthrough of~\cite{brunettietal2016} was the definition of field-dependent coordinates also for such a background, using the gauge-dependent parts of the metric perturbation to define corrections to the background coordinates.

Following~\cite{froeb2018}, we define functionals $X^{(\alpha)}$ in the full perturbed geometry as solutions of the scalar differential equation
\begin{equation}
\label{eqn_gen_harm_coords_ansatz}
\nabla^2 X^{(\alpha)} = \frac{1}{\sqrt{-g}} \partial_\mu \left( \sqrt{-g} g^{\mu\nu} \partial_\nu X^\alpha \right) = 0 \eqend{,}
\end{equation}
where $\nabla^2$ is the d'Alembert operator of the full metric $g$, $\alpha = 0,\ldots,n-1$, and we have enclosed the index $\alpha$ in parentheses to make it clear that the $X$ are scalar functionals and do not constitute a vector. By definition, the background value of these functionals are the Cartesian background coordinates $x^\alpha$, which are harmonic: $\partial^2 x^\alpha = 0$. We might thus call the $X^{(\alpha)}$ field-dependent or generalized harmonic coordinates. We then make the perturbative expansion
\begin{equation}
\label{eqn_gen_harm_coords_form_exp}
X^{(\alpha)} = x^\alpha + \sum_{k=1}^\infty \kappa^k X_{(k)}^{(\alpha)}(x) \eqend{,}
\end{equation}
and insert it together with the perturbative expansion of the metric~\eqref{eq_metric_expansion}, the inverse metric
\begin{equation}
\label{eq_inversemetric_expansion}
g^{\mu\nu} = \eta^{\mu\nu} - \kappa h^{\mu\nu} + \kappa^2 h^{\mu\rho} h^\nu{}_\rho + \bigo{\kappa^3}
\end{equation}
and the metric determinant
\begin{equation}
\label{eq_determinant_expansion}
\sqrt{-g} = 1 + \frac{1}{2} \kappa h + \frac{1}{8} \kappa^2 h^2 - \frac{1}{4} \kappa^2 h^{\mu\nu} h_{\mu\nu} + \bigo{\kappa^3}
\end{equation}
in equation~\eqref{eqn_gen_harm_coords_ansatz}. At first and second order in $\kappa$, this results in the following equations:
\begin{equations}[eqn_gen_harm_coords_order]
\partial^2 X_{(1)}^{(\alpha)} &= J_{(1)}^\alpha[h] \equiv \partial_\mu h^{\mu\alpha} - \frac{1}{2} \partial^\alpha h \eqend{,} \\
\partial^2 X_{(2)}^{(\alpha)} &= J_{(2)}^\alpha[h] + K_{(1)}[h] X_{(1)}^{(\alpha)} \eqend{,}
\end{equations}
where
\begin{equations}
J_{(2)}^\alpha[h] &\equiv \frac{1}{2} h_{\beta\gamma} \partial^\alpha h^{\beta\gamma} - h_{\beta\gamma} \partial^\beta h^{\gamma\alpha} - h^\alpha{}_\beta J_{(1)}^\beta[h] \eqend{,} \\ 
K_{(1)}[h] &\equiv h^{\alpha\beta} \partial_\alpha \partial_\beta + J_{(1)}^\gamma[h] \partial_\gamma \eqend{.}
\end{equations}

To ensure causality of the gauge-invariant observables~\cite{froeblima2018}, equations~\eqref{eqn_gen_harm_coords_order} must be solved using the retarded Green's function $G_\text{ret}$ of the Minkowski d'Alembert operator, which fulfills
\begin{equation}
\partial^2 G_\text{ret}(x,y) = \delta^n(x-y) \eqend{,} \qquad G_\text{ret}(x,y) = 0 \text{ for } (x-y)^2 > 0 \text{ or } x^0 < y^0 \eqend{.}
\end{equation}
While an explicit solution for $G_\text{ret}$ can easily be written down, we won't need it in the following. We thus obtain
\begin{equations}[eqn_field-dep_coords_expression]
X_{(1)}^{(\alpha)}(x) &= \int G_\text{ret}(x,y) J_{(1)}^\alpha(y) \total^n y \eqend{,} \\
X_{(2)}^{(\alpha)}(x) &= \int G_\text{ret}(x,y) \left[ J_{(2)}^\alpha(y) + K_{(1)}(y) X_{(1)}^{(\alpha)}(y) \right] \total^n y \eqend{,}
\end{equations}
and using the transformation of the metric perturbation under diffeomorphisms~\eqref{eq_metric_transform}, one straightforwardly computes
\begin{equations}[eq_coord_corrections_transform]
\begin{split}
\delta_\xi X_{(1)}^{(\alpha)}(x) &= \xi^\alpha(x) + \kappa \xi_\rho(x) h^{\rho\alpha}(x) \\
&\quad+ \kappa \int \xi_\rho(y) \partial^y_\mu G_\text{ret}(x,y) \left[ 2 \partial^{(\alpha} h^{\mu)\rho}(y) - \partial^\rho h^{\mu\alpha}(y) - \eta^{\alpha\mu} J_{(1)}^\rho(y) \right] \total^n y \eqend{,}
\end{split} \\
\begin{split}
\delta_\xi X_{(2)}^{(\alpha)}(x) &= \xi^\mu(x) \partial_\mu X_{(1)}^{(\alpha)}(x) - h^{\alpha\mu}(x) \xi_\mu(x) \\
&\quad+ \int \xi^\mu(y) \partial^y_\nu G_\text{ret}(x,y) \left[ \eta^{\alpha\nu} J^{(1)}_\mu(y) + \partial_\mu h^{\alpha\nu}(y) - 2 \partial^{(\alpha} h^{\nu)}{}_\mu(y) \right] \total^n y + \bigo{\kappa} \eqend{.}
\end{split}
\end{equations}
It follows that
\begin{equation}
\delta_\xi X^{(\alpha)} = \kappa \xi^\mu \partial_\mu X^{(\alpha)} + \bigo{\kappa^3} \eqend{,}
\end{equation}
i.e., these functionals do indeed transform as scalars under diffeomorphisms as they must.

To construct gauge-invariant relational observables we then have to evaluate the quantity of interest in the new coordinates $X^{(\alpha)}$. For this, we first need to invert the relation $X^{(\alpha)}(x)$~\eqref{eqn_gen_harm_coords_form_exp} to obtain
\begin{splitequation}
x^\alpha &= X^{(\alpha)} - \kappa X_{(1)}^{(\alpha)}(x) - \kappa^2 X_{(2)}^{(\alpha)}(x) + \bigo{\kappa^3} \\
&= X^{(\alpha)} - \kappa X_{(1)}^{(\alpha)}(X) + \kappa^2 X_{(1)}^{(\mu)}(X) \partial_\mu X_{(1)}^{(\alpha)}(X) - \kappa^2 X_{(2)}^{(\alpha)}(X) + \bigo{\kappa^3} \eqend{.}
\end{splitequation}
The invariant observable corresponding to the metric perturbation is then defined by performing a diffeomorphism of the metric $g_{\mu\nu}$ to the new coordinates $X^{(\alpha)}$ and subtracting the flat background metric:
\begin{splitequation}
\mathcal{H}_{\mu\nu}(X) &\equiv \frac{1}{\kappa} \left( \frac{\partial x^\rho(X)}{\partial X^\mu} \frac{\partial x^\sigma(X)}{\partial X^\nu} \tilde{g}_{\rho\sigma}(x(X)) - \eta_{\mu\nu} \right) \\
&= h_{\mu\nu}(X) - \eta_{\rho\mu} \partial_\nu X_{(1)}^{(\rho)}(X) - \eta_{\rho\nu} \partial_\mu X_{(1)}^{(\rho)}(X) + \bigo{\kappa} \eqend{.}
\end{splitequation}
Using the transformation of the metric perturbation~\eqref{eq_metric_transform} and the coordinate corrections~\eqref{eq_coord_corrections_transform} under diffeomorphisms, we obtain
\begin{equation}
\delta_\xi \mathcal{H}_{\mu\nu} = 2 \partial_{(\mu} \xi_{\nu)} - \eta_{\rho\mu} \partial_\nu \xi^\rho - \eta_{\rho\nu} \partial_\mu \xi^\rho + \bigo{\kappa} = \bigo{\kappa} \eqend{,}
\end{equation}
verifying to first order that our observable is invariant under diffeomorphisms. The expansion to higher orders is straightforward but lengthy. However, since $\mathcal{H}_{\mu\nu}$ is gauge-invariant, the computation can be done in any gauge for the metric perturbation and will be simplified by choosing a suitable gauge. In particular, we will choose the exact (Landau-type) gauge $J_{(1)}^\mu[h] = 0$ in subsection~\ref{sec_brst}, and can thus neglect all terms involving $J_{(1)}^\mu$ in the expansion. This is a significant simplification, and up to second order we only need
\begin{equation}
\label{eqn_gauge_inv_H_explicit}
\mathcal{H}_{\mu\nu} = h_{\mu\nu} - 2 \kappa \eta_{\rho(\mu} \partial_{\nu)} X_{(2)}^{(\rho)} + \bigo{\kappa^2} \eqend{,}
\end{equation}
where instead of equation~\eqref{eqn_field-dep_coords_expression} we can use
\begin{equation}
X_{(2)}^{(\alpha)}(x) = \int G_\text{ret}(x,y) \left[ \frac{1}{2} h_{\beta\gamma} \partial^\alpha h^{\beta\gamma} - h_{\beta\gamma} \partial^\beta h^{\gamma\alpha} \right](y) \total^n y \eqend{.}
\end{equation}

\subsection{Actions for gravity and a point particle}
\label{sec_action}

For the gravitational field, we employ the Einstein--Hilbert action
\begin{equation}
S_\text{G} = \frac{1}{\kappa^2} \int R \sqrt{-g} \total^n x \eqend{,}
\end{equation}
which in principle we need to expand to fourth order in metric perturbations. However, since the graviton is massless and massless tadpoles vanish in dimensional regularisation~\cite{leibbrandt1975}, we can ignore the four-graviton vertex which only leads to such tadpoles at one-loop order, and only the three-graviton vertex is relevant. Expressing the Ricci scalar in terms of Christoffel symbols and integrating by parts, we obtain
\begin{splitequation}
\label{eqn_einsteinhilbert}
S_\text{G} &= \frac{1}{4} \int g^{\mu\nu} g^{\alpha\beta} g^{\rho\sigma} \Big[ 2 \partial_\alpha h_{\mu\rho} \partial_\sigma h_{\nu\beta} - 2 \partial_\alpha h_{\mu\nu} \partial_\sigma h_{\rho\beta} - \partial_\alpha h_{\mu\rho} \partial_\beta h_{\nu\sigma} \\
&\qquad\qquad+ \partial_\alpha h_{\mu\nu} \partial_\beta h_{\rho\sigma} \Big] \sqrt{-g} \total^n x \eqend{,}
\end{splitequation}
from which the both the quadratic and cubic terms can be easily read off by expanding the inverse metric~\eqref{eq_inversemetric_expansion} and metric determinant~\eqref{eq_determinant_expansion} to first order. We refrain from giving the explicit expressions at this point, and collect them later in the full interaction.

The point particle is described by its worldline or position $z^\mu(\tau)$ and the action
\begin{equation}
\label{eqn_PPAction_general}
S_\text{PP} = \frac{1}{2} \int \left[ \frac{1}{e(\tau)} g_{\mu\nu}(z(\tau)) \dot{z}^\mu(\tau) \dot{z}^\nu(\tau) - m^2 e(\tau) \right] \total \tau \eqend{,}
\end{equation}
where a dot denotes a derivative with respect to the parameter $\tau$, $m$ is the mass of the particle, and the einbein $e(\tau)$ ensures reparametrization invariance of the action under arbitrary transformations $\tau \to \tau'(\tau)$. As is well known, $e$ is an auxiliary field and can be integrated out (i.e., replaced by its equation of motion), which leads to the square root form of the action, but the above form is more convenient for quantization.

Both the Einstein--Hilbert action~\eqref{eqn_einsteinhilbert} and the point particle action~\eqref{eqn_PPAction_general} are invariant under infinitesimal diffeomorphisms $\delta x^\mu = - \kappa \xi^\mu(x)$ that act on fields via the Lie derivative~\eqref{eq_metric_transform}:
\begin{equations}[eqn_Lie_derivative]
\mathcal{L}_\xi g_{\mu\nu} &= 2 \kappa \nabla_{(\mu} \xi_{\nu)} \eqend{,} \\
\mathcal{L}_\xi h_{\mu\nu} &= 2 \partial_{(\mu} \xi_{\nu)} + \kappa \xi^\rho \partial_\rho h_{\mu\nu} + 2 \kappa h_{\rho(\mu} \partial_{\nu)} \xi^\rho \eqend{,} \\
\mathcal{L}_\xi z^\mu(\tau) &= - \kappa \xi^\mu(z(\tau)) \eqend{,}
\end{equations}
while the point particle action is in addition invariant under worldline reparametrizations $\delta_\text{rep} \tau = - \epsilon(\tau)$, under which the fields transform according to
\begin{equations}[eqn_reparametrisation_transformations]
\delta_\text{rep} \, z^\mu(\tau) &= \epsilon(\tau) \dot{z}(\tau) \eqend{,}\\
\delta_\text{rep} \, g_{\mu\nu}(z(\tau)) &= \epsilon(\tau) \dot{z}^\alpha(\tau) \left( \partial_\alpha g_{\mu\nu} \right)(z(\tau)) \eqend{,}\\
\delta_\text{rep} \, e(\tau) &= \frac{\total}{\total \tau} \left[ \epsilon(\tau) e(\tau) \right] \eqend{.}
\end{equations}
Since the metric in the point particle action is evaluated on the worldline of the particle, it also changes under these transformations, in addition to its change under diffeomorphisms.

To preserve the above invariance at each order in perturbation theory, it is necessary to also perturbatively expand the particle's position and the einbein. On the background, we choose the particle to be at rest at the origin, which results in a constant and purely time-like four-velocity
\begin{equation}
\label{eq_vmu_def}
v^\mu \equiv \dot z_0^\mu = (1,\vec{0})^\mu \eqend{.}
\end{equation}
Moreover, the equation of motion for the einbein determines its background value to be $e_0(\tau) = \sqrt{ - \eta_{\mu\nu} \dot{z}_0^\mu(\tau) \dot{z}_0^\nu(\tau) }/m = 1/m$. Since we are interested in the corrections to the Newtonian potential sourced by a graviton loop, we furthermore suppress the fluctuations of the particle itself by using a large mass expansion, and set
\begin{equation}
\label{eq_ze_largemassexpansion}
z^\mu(\tau) = v^\mu \tau + \frac{1}{\sqrt{m}} y^\mu(\tau) \eqend{,} \qquad e(\tau) = \frac{1}{m} \left[ 1 + \frac{1}{\sqrt{m}} f(\tau) \right] \eqend{.}
\end{equation}
However, we will see that a certain amount of fluctuations of the particle survive even for large $m$, and can be interpreted as analogous to an intrinsically quantum-mechanical ``Zitterbewegung''~\cite{schroedinger1930,thaller}. As the ``Zitterbewegung'' pertains to Fermi fields which are not modelled here, this is nevertheless by analogy only. It is an expression of the quantum fluctuations of the position and energy-momentum of the particle and their backreaction via the gravitational interaction. The ``Zitterbewegung'' has also been investigated in gravitational and non-commutative spacetime contexts~\cite{kobakhidzemanningtureanu2016,ecksteinfrancomiller2017}. For the point particle action, we then obtain up to an irrelevant constant and integration by parts
\begin{splitequation}
\label{eq_SPP_expanded}
S_\text{PP} &= \frac{1}{2} \int \bigg[ - f^2(\tau) - 2 f(\tau) v_\mu \dot y^\mu(\tau) + \eta_{\mu\nu} \dot y^\mu(\tau) \dot y^\nu(\tau) \\
&\qquad+ \frac{1}{\sqrt{m}} \left[ f^3(\tau) + 2 f^2(\tau) v_\mu \dot y^\mu(\tau) - f(\tau) \eta_{\mu\nu} \dot y^\mu(\tau) \dot y^\nu(\tau) \right] \\
&\qquad+ \kappa \left[ m - \sqrt{m} f(\tau) + f(\tau)^2 - \frac{1}{\sqrt{m}} f^3(\tau) \right] h_{\mu\nu}(z(\tau)) v^\mu v^\nu \\
&\qquad+ 2 \kappa \left( \sqrt{m} - f(\tau) + \frac{1}{\sqrt{m}} f^2(\tau) \right) h_{\mu\nu}(z(\tau)) v^\nu \dot y^\mu(\tau) \\
&\qquad+ \kappa \left[ 1 - \frac{1}{\sqrt{m}} f(\tau) \right] h_{\mu\nu}(z(\tau)) \dot y^\mu(\tau) \dot y^\nu(\tau) \bigg] \total \tau + \bigo{m^{-1}}
\end{splitequation}
with
\begin{splitequation}
\label{eq_hmunu_ztau}
h_{\mu\nu}(z(\tau)) &= h_{\mu\nu}(z_0(\tau)) + \frac{1}{\sqrt{m}} y^\alpha(\tau) \partial_\alpha h_{\mu\nu}(z_0(\tau)) + \frac{1}{2 m} y^\alpha(\tau) y^\beta(\tau) \partial_\alpha \partial_\beta h_{\mu\nu}(z_0(\tau)) \\
&\quad+ \frac{1}{6 m^\frac{3}{2}} y^\alpha(\tau) y^\beta(\tau) y^\gamma(\tau) \partial_\alpha \partial_\beta \partial_\gamma h_{\mu\nu}(z_0(\tau)) \eqend{,}
\end{splitequation}
and the power of $m^{-1/2}$ in the expansion~\eqref{eq_ze_largemassexpansion} was chosen such that all fields have canonical kinetic terms, given in the first line of~\eqref{eq_SPP_expanded}. Note that the lowest-order interaction term can be written as
\begin{equation}
\label{eq_pp_hmnunu_tmunu_vertex}
\frac{\kappa m}{2} \int h_{\mu\nu}(z_0(\tau)) v^\mu v^\nu \total \tau = \frac{\kappa}{2} \int h_{\mu\nu}(x) T_0^{\mu\nu}(x) \total^n x
\end{equation}
with the point particle's stress tensor
\begin{equation}
\label{eq_pp_stresstensor}
T_0^{\mu\nu}(x) = m v^\mu v^\nu \delta^{n-1}(\vec{x}) \eqend{.}
\end{equation}

\subsection{The BRST method: gauge-fixing and propagators}
\label{sec_brst}

As we have seen in the last subsection, the gauge symmetries in our problem are infinitesimal diffeomorphisms and worldline reparametrizations, which both need to be gauge-fixed for quantization, and we employ the BRST formalism~\cite{becchirouetstora1975,becchirouetstora1976,kugoojima1978,weinberg_v2,barnichbrandthenneaux2000} to do so. The central object is a fermionic nilpotent differential $\brst$, which acts on fields like a symmetry transformation with the parameter replaced by a new ghost field, which has opposite statistics to the symmetry parameter. Furthermore, for each ghost one introduces an antighost and an auxiliary field, which are needed for the gauge fixing of the corresponding symmetry. We therefore introduce a ghost field $c_\mu$, an antighost $\bar{c}_\mu$ and the auxiliary field $B_\mu$ for the diffeomorphisms and a ghost field $c$, an antighost $\bar{c}$ and an auxiliary field $b$ for the reparametrizations. Ghosts and antighosts are fermions, while the auxiliary fields are bosonic. We then postulate the transformations
\begin{equations}[eqn_BRST_op_def]
\brst h_{\mu\nu}(x) &= 2 \partial_{(\mu} c_{\nu)}(x) + \kappa c^\rho(x) \partial_\rho h_{\mu\nu}(x) + 2 \kappa h_{\rho(\mu}(x) \partial_{\nu)} c^\rho(x) \eqend{,} \\
\brst c_\mu(x) &= \kappa c^\rho(x) \partial_\rho c_\mu(x) \eqend{,} \\ 
\brst \bar{c}_\mu(x) &= B_\mu(x) \eqend{,} \\ 
\brst B_\mu(x) &= 0 \eqend{,} \\
\begin{split}
\brst y^\mu(\tau) &= c(\tau) \left[ v^\mu + \frac{1}{\sqrt{m}} \dot y^\mu(\tau) \right] - \kappa \sqrt{m} c^\mu(z_0(\tau)) - \kappa y^\alpha(\tau) \partial_\alpha c^\mu(z_0(\tau)) \\
&\quad- \frac{\kappa}{2 \sqrt{m}} y^\alpha(\tau) y^\beta(\tau) \partial_\alpha \partial_\beta c^\mu(z_0(\tau)) + \bigo{ \frac{1}{m} } \eqend{,}
\end{split} \\
\brst f(\tau) &= \dot c(\tau) + \frac{1}{\sqrt{m}} \frac{\total}{\total \tau} \left[ c(\tau) f(\tau) \right] \eqend{,} \\
\brst c(\tau) &= \frac{1}{\sqrt{m}} c(\tau) \dot c(\tau) \eqend{,} \\
\brst \bar{c}(\tau) &= b(\tau) \eqend{,} \\
\brst b(\tau) &= 0 \eqend{,}
\end{equations}
which for the fields $h_{\mu\nu}$, $y^\mu$ and $f$ follows directly from the transformations~\eqref{eqn_Lie_derivative} and~\eqref{eqn_reparametrisation_transformations} together with the expansion~\eqref{eq_ze_largemassexpansion}. On the ghosts, the action of $\brst$ is fixed by requiring nilpotency $\brst^2 = 0$ on the fields, while the action on the antighosts and auxiliary fields is the standard one. 

With these definitions and a graded Leibniz rule for $\brst$, it follows that $\brst^2 = 0$ when acting on any local polynomial of fields, ghosts and other fields, and the gauge invariance of the gravitational~\eqref{eqn_einsteinhilbert} and point-particle action~\eqref{eq_SPP_expanded} is replaced by BRST symmetry: $\brst \left( S_\text{G} + S_\text{PP} \right) = 0$. Due to the nilpotency of the BRST differential, we can add to the action any BRST-exact term, from which we obtain the gauge-fixing and ghost actions. Concretely, we add the following terms:
\begin{equation}
\brst \int \bar{c}^\mu \left( \frac{\zeta}{2} B_\mu + \partial^\nu h_{\mu\nu} - \frac{1}{2} \partial_\mu h \right) \total^n x - \brst \int \bar{c} \left( \frac{1}{2} b + \xi v_\mu \dot y^\mu + \frac{1}{\xi} f \right) \total \tau \eqend{,}
\end{equation}
where $\zeta$ and $\xi$ are real constants that we want to send to zero once the propagators have been determined, to impose the exact (Landau-type) gauge that simplifies the computation. Performing the BRST transformations, we obtain the gauge-fixing action (terms quadratic in the fields and auxiliary fields, no ghosts or antighosts), the free ghost action (terms quadratic in ghosts and antighosts) and additional interaction terms (proportional to $\kappa$ and/or inverse powers of $\sqrt{m}$). The gauge-fixing action can be combined with the free part of the gravitational and point-particle action to give
\begin{splitequation}
S_{0,\text{GPP}} &\equiv \frac{1}{2} \int \left( h^{\mu\nu} P_{\mu\nu\rho\sigma} h^{\rho\sigma} + \zeta \tilde{B}^\mu \tilde{B}_\mu \right) \total^n x \\
&- \frac{1}{2} \int \left[ \left( 1 - \frac{1}{\xi^2} \right) f^2(\tau) + y^\mu(\tau) \left( \eta_{\mu\nu} + \xi^2 v_\mu v_\nu \right) \partial_\tau^2 y^\nu(\tau) + \tilde{b}^2 \right] \total \tau \eqend{,}
\end{splitequation}
with
\begin{equation}
\tilde{B}_\mu = B_\mu + \frac{1}{\zeta} \left( \partial^\nu h_{\mu\nu} - \frac{1}{2} \partial_\mu h \right) \eqend{,} \qquad \tilde{b} = b + \xi v_\mu \dot y^\mu + \frac{1}{\xi} f
\end{equation}
and the symmetric differential operator
\begin{splitequation}
P_{\mu\nu\rho\sigma} &\equiv \frac{1}{2} \eta_{\mu(\rho} \eta_{\sigma)\nu} \partial^2 - \left( 1 - \frac{1}{\zeta} \right) \partial_{(\mu} \eta_{\nu)(\rho} \partial_{\sigma)} \\
&\quad+ \left( \frac{1}{2} - \frac{1}{2 \zeta} \right) \left( \eta_{\mu\nu} \partial_\rho \partial_\sigma + \eta_{\rho\sigma} \partial_\mu \partial_\nu \right) - \left( \frac{1}{2} - \frac{1}{4\zeta} \right) \eta_{\mu\nu} \eta_{\rho\sigma} \partial^2 \eqend{,}
\end{splitequation}
from which the propagators are obtained by inverting the differential operators. The ghost action reads
\begin{equation}
S_{0,\text{GH}} \equiv - \int \bar{c}^\mu \partial^2 c_\mu \total^n x + \left( \frac{1}{\xi} - \xi \right) \int \bar{c} \dot c \total \tau \eqend{,}
\end{equation}
and the additional interactions are given by
\begin{splitequation}
\label{eq_SintGH_full}
S_\text{I,GH} &\equiv \kappa \int \left( \partial^\mu \bar{c}^\nu - \frac{1}{2} \eta^{\mu\nu} \partial_\sigma \bar{c}^\sigma \right) \left( c^\rho \partial_\rho h_{\mu\nu} + 2 h_{\rho(\mu} \partial_{\nu)} c^\rho \right) \total^n x \\
&\quad+ \kappa \xi \int v_\mu \dot {\bar{c}} \left[ \sqrt{m} c^\mu(z_0) + y^\alpha \partial_\alpha c^\mu(z_0) + \frac{1}{2 \sqrt{m}} y^\alpha y^\beta \partial_\alpha \partial_\beta c^\mu(z_0) \right] \total \tau \\
&\quad- \frac{1}{\sqrt{m}} \int \dot {\bar{c}} c \left( \xi v_\mu \dot y^\mu + \frac{1}{\xi} f \right) \total \tau + \bigo{ \frac{1}{m} } \eqend{.}
\end{splitequation}

Since the shifted auxiliary fields $\tilde{B}_\mu$ and $\tilde{b}$ decouple from the rest of the fields, they do not contribute to any expectation values. For the rest of the fields and the ghosts, we obtain the propagators $G^{AB}(x,x') = - \mathi \expect{ \mathcal{T} A(x) B(x') }$ by inverting the differential operators appearing in the free action, which results in
\begin{equations}[eq_propagators_full]
G^{hh}_{\mu\nu\rho\sigma}(x,x') &= \left( 2 \eta_{\mu(\rho} \eta_{\sigma)\nu} - \frac{2}{n-2} \eta_{\mu\nu} \eta_{\rho\sigma} - 4 (1-\zeta) \frac{\partial_{(\mu} \eta_{\nu)(\rho} \partial_{\sigma)}}{\partial^2} \right) G(x,x') \eqend{,} \\
G^{c\bar{c}}_{\mu\nu}(x,x') &= - \eta_{\mu\nu} G(x,x') \eqend{,} \\
G^{ff}(\tau,\tau') &= \frac{\xi^2}{1 - \xi^2} \delta(\tau-\tau') \eqend{,} \\
G_{yy}^{\mu\nu}(\tau,\tau') &= - \frac{1}{2} \left( \eta^{\mu\nu} - \frac{\xi^2}{1 - \xi^2} v^\mu v^\nu \right) \abs{\tau-\tau'} \eqend{,} \\
G^{c\bar{c}}(\tau,\tau') &= \frac{\xi}{2 (1-\xi^2)} \operatorname{sgn}(\tau-\tau') \eqend{,}
\end{equations}
where $G(x,x')$ is the massless flat space Feynman propagator given by
\begin{equation}
G(x,x') = - \int \frac{\mathe^{\mathi p(x-x')}}{p^2 - \mathi \epsilon} \frac{\total^n p}{(2\pi)^n} \eqend{,} \qquad \partial^2 G(x,x') = \delta^n(x-x') \eqend{.}
\end{equation}
As $\xi \to 0$, the propagators for $f$ and the reparametrization ghost $c$ vanish and the ones for $h_{\mu\nu}$ and $y^\mu$ simplify. However, there is an interaction in~\eqref{eq_SintGH_full} which diverges as $\xi \to 0$, so we have to be careful in taking this limit. However, this interaction couples the reparametrization ghost $c$ and antighost $\bar{c}$ with the einbein perturbation $f$, and since the propagator for $f$ vanishes quadratically with $\xi$ and the one for $c$ and $\bar{c}$ linearly, the interaction vertex together with half of each propagator vanishes like $\xi^{-1} \sqrt{\xi^2} \sqrt{\xi} \sqrt{\xi} \sim \xi$, such that any Feynman diagram involving this vertex will vanish. It follows that we can take the limit $\xi \to 0$ with impunity, and in particular neither the einbein perturbation nor the reparametrization ghost contribute to expectation values which greatly reduces the number of diagrams that we have to compute.

\section{Perturbative corrections, quantum and classical}
\label{sec_correction}

We are now prepared to compute the expectation value of the gauge-invariant observable corresponding to the metric perturbation to one-loop order. As usual, this is done using the Gell-Mann--Low formula
\begin{equation}
\expect{ \mathcal{H}_{\mu\nu}(x) } = \frac{\expect{ \mathcal{H}_{\mu\nu}(x) \mathe^{\mathi S_\text{int}} }_0}{\expect{ \mathe^{\mathi S_\text{int}} }_0} \eqend{,}
\end{equation}
where the free expectation value is computed by Wick's formula with the propagators~\eqref{eq_propagators_full}. Since neither the einbein perturbation nor the reparametrization ghost contribute in the gauge we are using, only the following propagators are relevant:
\begin{equations}[eq_propagators]
G^{hh}_{\mu\nu\rho\sigma}(x,x') &= \left( 2 \eta_{\mu(\rho} \eta_{\sigma)\nu} - \frac{2}{n-2} \eta_{\mu\nu} \eta_{\rho\sigma} - 4 \frac{\partial_{(\mu} \eta_{\nu)(\rho} \partial_{\sigma)}}{\partial^2} \right) G(x,x') \eqend{,} \\
G^{c\bar{c}}_{\mu\nu}(x,x') &= - \eta_{\mu\nu} G(x,x') \eqend{,} \\
G_{yy}^{\mu\nu}(\tau,\tau') &= - \frac{1}{2} \eta^{\mu\nu} \abs{\tau-\tau'} \eqend{.}
\end{equations}
The interaction $S_\text{int}$ is obtained from equations~\eqref{eqn_einsteinhilbert}, \eqref{eq_SPP_expanded} and~\eqref{eq_SintGH_full}, and the terms relevant for the one-loop computation read
\begin{splitequation}
\label{eq_sint}
S_\text{int} &= \frac{\kappa}{4} \int \left( \frac{1}{2} \eta^{\mu\nu} \eta^{\alpha\beta} \eta^{\rho\sigma} h - \eta^{\mu\nu} \eta^{\alpha\beta} h^{\rho\sigma} - \eta^{\mu\nu} \eta^{\rho\sigma} h^{\alpha\beta} - \eta^{\alpha\beta} \eta^{\rho\sigma} h^{\mu\nu} \right) \\
&\qquad\quad\times \Big( 2 \partial_\alpha h_{\mu\rho} \partial_\sigma h_{\nu\beta} - 2 \partial_\alpha h_{\mu\nu} \partial_\sigma h_{\rho\beta} - \partial_\alpha h_{\mu\rho} \partial_\beta h_{\nu\sigma} + \partial_\alpha h_{\mu\nu} \partial_\beta h_{\rho\sigma} \Big) \total^n x \\
&\quad+ \kappa \int \left( \partial^\mu \bar{c}^\nu - \frac{1}{2} \eta^{\mu\nu} \partial_\sigma \bar{c}^\sigma \right) \left( c^\rho \partial_\rho h_{\mu\nu} + 2 h_{\rho(\mu} \partial_{\nu)} c^\rho \right) \total^n x \\
&\quad+ \frac{\kappa}{2} \int \Big[ m \, h_{\mu\nu}(z(\tau)) v^\mu v^\nu + 2 \sqrt{m} \, h_{\mu\nu}(z(\tau)) v^\nu \dot y^\mu(\tau) + h_{\mu\nu}(z(\tau)) \dot y^\mu(\tau) \dot y^\nu(\tau) \Big] \total \tau \eqend{,}
\end{splitequation}
where $h_{\mu\nu}(z(\tau))$ is given in equation~\eqref{eq_hmunu_ztau}.
We see that there are two different types of interactions: the first are the well-known ones that couple gravitons or gravitons and diffeomorphism ghosts, while the second ones couples the graviton with worldline perturbations. Moreover, we have also contributions coming from the coordinate corrections to $\mathcal{H}_{\mu\nu}$~\eqref{eqn_gauge_inv_H_explicit}. Working to order $\kappa^4 \sim G_\text{N}^2$, the graviton and ghost loop Feynman diagrams arising from the first type of interaction are computed in section~\ref{sec_loop}, the diagrams involving worldline perturbations are computed in section~\ref{sec_worldline}, and the corrections from the field-dependent coordinates are computed in section~\ref{sec_coord}. There are also classical corrections which are given in section~\ref{sec_class}, and moreover many diagrams actually turn out to vanish, which we list for completeness (together with the reason for their vanishing) in section~\ref{sec_vanish}.

\subsection{Graviton and ghost loop corrections}
\label{sec_loop}

\begin{figure}[ht]
\begin{center}
\begin{fmffile}{grav_ghost_loop}
    \begin{fmfgraph*}(.4,.15)
        \fmfdotn{v}{2}
        \fmfleft{hin1}
        \fmfright{hout1}
        \fmfv{decoration.shape=circle,decoration.filled=30,decor.size=3.5mm}{hout1}
        \fmf{wiggly}{hin1,v1}
        \fmf{wiggly}{v2,hout1}
        \fmf{wiggly,right}{v1,v2}
        \fmf{wiggly,right}{v2,v1}
        \fmfforce{(.3w,.5h)}{v1}
        \fmfforce{(.7w,.5h)}{v2}
        \fmfv{label=\small{$ \kappa$},label.angle=135}{v1}
        \fmfv{label=\small{$ \kappa$},label.angle=45}{v2}
        \fmfv{label=\small{$ m\kappa$},label.angle=90,label.dist=3mm}{hout1}
        \fmfv{label=\small{$ \kappa$}, label.angle=90}{hin1}
    \end{fmfgraph*}
    \hspace{.1\linewidth}
    \begin{fmfgraph*}(.4,.15)
        \fmfdotn{v}{2}
        \fmfleft{hin1}
        \fmfright{hout1}
        \fmfv{decoration.shape=circle,decoration.filled=30,decor.size=3.5mm}{hout1}
        \fmf{wiggly}{hin1,v1}
        \fmf{wiggly}{v2,hout1}
        \fmf{ghost,right}{v1,v2}
        \fmf{ghost,right}{v2,v1}
        \fmfforce{(.3w,.5h)}{v1}
        \fmfforce{(.7w,.5h)}{v2}
        \fmfv{label=\small{$ \kappa$},label.angle=135}{v1}
        \fmfv{label=\small{$ \kappa$},label.angle=45}{v2}
        \fmfv{label=\small{$ m\kappa$},label.angle=90,label.dist=3mm}{hout1}
        \fmfv{label= \small{$ \kappa$}, label.angle=90}{hin1}
    \end{fmfgraph*}
	\vspace{.02\linewidth}
\end{fmffile}
\caption{The graviton and ghost loop contributions to $\protect\expect{ \mathcal{H}_{\mu\nu}(x) }$, where wiggly lines denote graviton propagators, dotted lines denote diffeomorphism ghost propagators, the point particle's stress tensor is shown with a grey circle, and the observation point $x$ is at the left of each diagram. With the proper factors of $\kappa$ shown in the diagrams, both give quantum corrections proportional to $m \kappa^4 \propto G_\text{N}^2 m$ to the gravitational potential.}
\label{fig_grav_ghost_loop}
\end{center}
\end{figure}
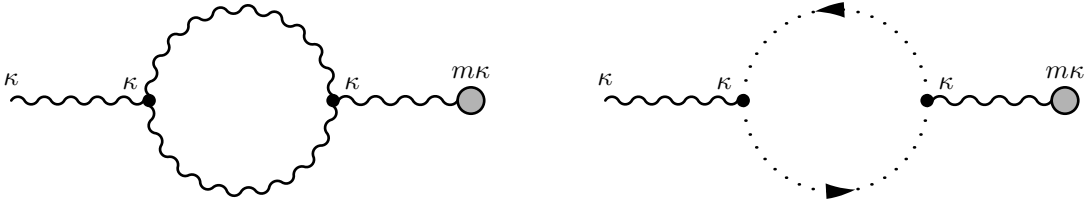

The computation of the one-loop corrections due to gravitons and diffeomorphism ghosts is straightforward, and parts of the computation (namely the graviton self-energy) have been performed numerous times before, see for example~\cite{cappernamazie1978,capper1980,hsukim2012}. To obtain from the graviton self-energy the relevant one-loop corrections to our invariant observable, we have to attach a free graviton propagator connecting the loop with the observation point $x$ at which the observable is evaluated, and another graviton propagator connecting it with the point particle's stress tensor, as depicted in Fig.~\ref{fig_grav_ghost_loop}.
Since the computation of the graviton self-energy $\Sigma_{\mu\nu\rho\sigma}(p)$ is standard, we will not give any details here, and merely note that we agree with the results of~\cite{cappernamazie1978,capper1980} after correcting for the different conventions and some typos as noted in~\cite{hsukim2012}. Attaching the graviton propagators and point particle stress tensor~\eqref{eq_pp_stresstensor} then results in the expression
\begin{splitequation}
&\frac{m \kappa^4}{2} \iiint G^{hh}_{\mu\nu\rho\sigma}(x,y) \int \Sigma^{\rho\sigma\alpha\beta}(p) \mathe^{\mathi p (y-y')} \frac{\total^n p}{(2\pi)^n} G^{hh}_{\alpha\beta\gamma\delta}(y',z_0(\tau)) v^\gamma v^\delta \total^n y \total^n y' \total \tau \\
&= \frac{m \kappa^4}{2} \int \int \tilde{G}^{hh}_{\mu\nu\rho\sigma}(p) \Sigma^{\rho\sigma\alpha\beta}(p) \tilde{G}^{hh}_{\alpha\beta\gamma\delta}(p) \mathe^{\mathi p (x-z_0(\tau))} v^\gamma v^\delta \total \tau \frac{\total^n p}{(2\pi)^n} 
\end{splitequation}
with the Fourier transform $\tilde{G}^{hh}$ of the graviton propagator. Since $z_0^\mu(\tau) = v^\mu \tau$ with constant $v^\mu$~\eqref{eq_vmu_def}, the integral over $\tau$ results in $(2\pi) \delta(p^0)$ and thus a vanishing time component of the four-momentum. The tensor algebra is performed with the help of the tensor algebra package xAct~\cite{xact}, and spatial momenta $\vec{p}_\mu$ are converted into derivatives with respect to $\vec{x}^\mu$. The remaining integral over the spatial components $\vec{p}$ can be done with the help of the well-known formula
\begin{equation}
\label{eq_p2_fouriertrafo}
\int \left( \vec{p}^2 \right)^a \mathe^{\mathi \vec{p} \vec{x}} \frac{\total^{n-1} \vec{p}}{(2 \pi)^{n-1}} = \frac{4^a \Gamma\left( a + \frac{n-1}{2} \right)}{\pi^\frac{n-1}{2} \Gamma(-a)} \left( \vec{x}^2 \right)^{- a - \frac{n-1}{2}}
\end{equation}
(see for example~\cite{smirnov2004}, with the simplification that no Wick rotation is necessary). The final result for the contribution from the graviton and ghost loops to the expectation value of our invariant observable $\expect{ \mathcal{H}_{\mu\nu}(x) }$ is then given by
\begin{splitequation}
\label{eq_corr_loop_pre}
&\frac{164 G_\text{N}^2 m}{15 \pi r^3} \left( 3 \frac{\vec{x}_\mu \vec{x}_\nu}{r^2} - \bar\eta_{\mu\nu} \right) \left( - \frac{1}{n-4} + \gamma + \ln(4 \pi) + \ln \mu^2 \right) \\
&\quad+ \frac{164 G_\text{N}^2 m}{15 \pi r^3} \left( 3 \frac{\vec{x}_\mu \vec{x}_\nu}{r^2} - \bar\eta_{\mu\nu} \right) \ln \left( \frac{r^2}{\mu^2} \right) \\
&\quad+ \frac{41 G_\text{N}^2 m}{5 \pi r^3} v_\mu v_\nu - \frac{4408 G_\text{N}^2 m}{75 \pi r^3} \frac{\vec{x}_\mu \vec{x}_\nu}{r^2} + \frac{2153 G_\text{N}^2 m}{225 \pi r^3} \bar\eta_{\mu\nu} \eqend{,}
\end{splitequation}
where $\mu$ is the renormalization scale, $r \equiv \abs{\vec{x}}$, and where we defined the purely spatial metric
\begin{equation}
\label{eq_bareta_def}
\bar{\eta}_{\mu\nu} \equiv \eta_{\mu\nu} + v_\mu v_\nu \eqend{.}
\end{equation}
To renormalize, we note that one can add a homogeneous solution to the generalized harmonic coordinates~\eqref{eqn_gen_harm_coords_ansatz}, i.e., a field-independent term that solves the wave equation, which is similar to the well-known residual gauge freedom in harmonic gauge. In particular, we have $\partial^2 \left( \vec{x}^\mu \, r^{1-n} \right) = 0$, and can thus add $c \, \vec{x}^\mu \, r^{1-n}$ with an arbitrary constant $c$ to the $X^{(\mu)}$, which gives an additional contribution $- 2 c \, \eta_{\rho(\mu} \partial_{\nu)} \left( \vec{x}^\rho \, r^{1-n} \right)$ to the expectation value of our invariant observable $\expect{ \mathcal{H}_{\mu\nu}(x) }$. Choosing
\begin{equation}
c = - \frac{82}{15} \frac{G_\text{N}^2 m}{\pi} \left( - \frac{1}{n-4} + \gamma + \ln(4 \pi) + \ln \mu - \frac{299}{205} \right) + \bigo{n-4} \eqend{,}
\end{equation}
we cancel the divergent part and obtain a finite result for the contribution from the graviton and ghost loops:
\begin{splitequation}
\label{eq_corr_loop}
\frac{82 G_\text{N}^2 m}{15 \pi r^3} \left( 3 \frac{\vec{x}_\mu \vec{x}_\nu}{r^2} - \bar\eta_{\mu\nu} \right) \ln \left( \frac{r^2}{\mu^2} \right) + \frac{41 G_\text{N}^2 m}{5 \pi r^3} v_\mu v_\nu - \frac{287 G_\text{N}^2 m}{45 \pi r^3} \bar\eta_{\mu\nu} \eqend{.}
\end{splitequation}
Similar methods have been used in the context of the post-Newtonian treatment of the motion of binary systems~\cite{blanchetdamourespositofarese2004,blanchetetal2005,jaranowskischaefer2013,bernardetal2015}, where the divergences that appear from treating point sources in classical General Relativity also could be absorbed in a redefinition of coordinates, using the residual gauge freedom of harmonic coordinates.

Let us stress here that compared to most previous works that use Feynman gauge (also called de Donder gauge), we are using a different gauge (generalised Landau gauge), where the graviton propagator is given by~\eqref{eq_propagators}. This has the advantage that the number of diagrams that we need to compute is reduced quite dramatically: for example, all diagrams that would involve the field-dependent coordinate correction $X^\mu_{(1)}$ vanish. (The remaining coordinate corrections involving $X^\mu_{(2)}$ are computed in section~\ref{sec_coord}.) Since individual Feynman diagrams are gauge-dependent, a direct comparison of individual diagrams with previous works is not possible; only the gauge-invariant end result can be compared, which we do in section~\ref{sec_results}.

\subsection{Worldline corrections}
\label{sec_worldline}

\begin{figure}[ht]
\begin{center}
\vspace{0.1em}
\begin{fmffile}{worldline_loop}
    \begin{fmfgraph*}(.4,.15)
        \fmfdotn{v}{1}
        \fmfleft{hin1}
        \fmfright{hout1,hout2}
        \fmf{wiggly}{hin1,v1}
        \fmf{wiggly}{v1,hout2}
        \fmf{wiggly}{v1,hout1}
        \fmf{plain}{hout2,hout1}
        \fmfv{decoration.shape=circle,decoration.filled=full,decoration.size=1.5mm}{hout1,hout2}
        \fmfforce{(.4w,.5h)}{v1}
        \fmfv{label= \small{$ \sqrt{m} \kappa$}}{hout2}
        \fmfv{label= \small{$ \sqrt{m} \kappa$}}{hout1}
        \fmfv{label= \small{$ \kappa$},label.angle=90}{hin1}
        \fmfv{label= \small{$ \kappa$},label.dist=3mm,label.angle=135}{v1}
    \end{fmfgraph*}
	\vspace{.02\linewidth}
\end{fmffile}
\caption{The wordline loop contribution to $\protect\expect{ \mathcal{H}_{\mu\nu}(x) }$, where a straight line denotes the propagator for the perturbed worldline. With the proper factors of $\kappa$ shown in the diagram, we again obtain a quantum correction proportional to $m \kappa^4 \propto G_\text{N}^2 m$ to the gravitational potential.}
\label{fig_worldline_loop}
\end{center}
\end{figure}
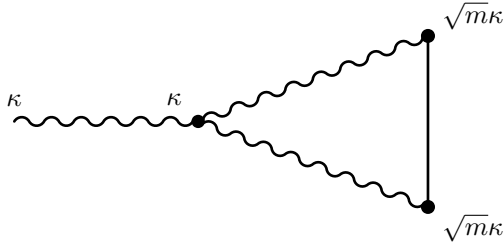

In contrast to the corrections from graviton and ghost loops, the corrections due to the fluctuating worldline of the point particle are completely new and have so far not been computed at all. The relevant graviton-worldline interaction vertices are given in the last line of~\eqref{eq_sint}. At one-loop order, there is one non-vanishing diagram depicted in Fig.~\ref{fig_worldline_loop}, where in contrast to the three-graviton vertex the graviton-worldline interaction only involves one graviton $h_{\mu\nu}$ and one worldline perturbation $y^\mu$.

The three-graviton vertex on the left side of the diagram in Fig.~\ref{fig_worldline_loop} is again well-known and can be read off from~\eqref{eq_sint}, and we focus on the right side involving the worldline propagator. The graviton-worldline-graviton interaction results in the expression
\begin{splitequation}
&\frac{\kappa \sqrt{m}}{2} \iint \Bigg[ \partial_\kappa G^{hh}_{\alpha\beta\mu\nu}(y, z_0(\tau)) \partial_\lambda G^{hh}_{\gamma\delta\rho\sigma}(y', z_0(\tau')) v^\mu v^\nu v^\rho v^\sigma G_{yy}^{\kappa\lambda}(\tau, \tau') \\
&\qquad\qquad- 2 \partial_\kappa G^{hh}_{\alpha\beta\mu\nu}(y, z_0(\tau)) G^{hh}_{\gamma\delta\rho\sigma}(y', z_0(\tau')) v^\mu v^\nu v^\sigma \partial_{\tau'} G_{yy}^{\kappa\rho}(\tau, \tau') \\
&\qquad\qquad- 2 G^{hh}_{\alpha\beta\mu\nu}(y, z_0(\tau)) \partial_\lambda G^{hh}_{\gamma\delta\rho\sigma}(y', z_0(\tau')) v^\nu v^\rho v^\sigma \partial_\tau G_{yy}^{\mu\lambda}(\tau, \tau') \\
&\qquad\qquad+ 4 G^{hh}_{\alpha\beta\mu\nu}(y, z_0(\tau)) G^{hh}_{\gamma\delta\rho\sigma}(y', z_0(\tau')) v^\nu v^\sigma \partial_\tau \partial_{\tau'} G_{yy}^{\mu\rho}(\tau, \tau') \Bigg] \total \tau \total \tau' \eqend{,}
\end{splitequation}
where the external gravitons (the ones that enter in the three-graviton vertex from the right) are $h_{\alpha\beta}(y)$ and $h_{\gamma\delta}(y')$. Integrating the $\tau$ derivatives by parts and transforming to Fourier space, this results in
\begin{splitequation}
&\frac{\kappa \sqrt{m}}{2} \iint \tilde{G}^{hh}_{\alpha\beta\mu\nu}(p) \tilde{G}^{hh}_{\gamma\delta\rho\sigma}(q) p_\kappa q_\lambda v^\nu v^\sigma \left[ \frac{1}{2} v^\mu v^\rho \eta^{\kappa\lambda} - v^\mu v^\lambda \eta^{\kappa\rho} - v^\rho v^\kappa \eta^{\mu\lambda} + 2 v^\kappa v^\lambda \eta^{\mu\rho} \right] \\
&\qquad\times \abs{\tau - \tau'} \mathe^{\mathi p (y-z_0(\tau)) + \mathi q (y'-z_0(\tau'))} \total \tau \total \tau' \frac{\total^n p}{(2\pi)^n} \frac{\total^n q}{(2\pi)^n} \eqend{.}
\end{splitequation}
To perform the worldline parameter integrals, we first shift $p \to p-q$ and $\tau \to \tau + \tau'$, and the integral over $\tau'$ results as before in $(2\pi) \delta(p^0)$. For the remaining integral, it turns out to be easier to first add the three-graviton vertex and the remaining graviton propagator, which (with the shift $p \to p-q$ that we performed) has momentum $p$. We introduce Feynman parameters according to
\begin{equation}
\frac{1}{[ (p-q)^2 - \mathi \epsilon ]^a ( q^2 - \mathi \epsilon )^b} = \frac{\Gamma(a+b)}{\Gamma(a) \Gamma(b)} \int_0^1 \frac{\xi^{a-1} (1-\xi)^{b-1}}{\left[ \vec{q}^2 - (q^0)^2 + \left[ \left( \vec{p} - \vec{q} \right)^2 - \vec{q}^2 \right] \xi - \mathi \epsilon \right]^{a+b}} \total \xi \eqend{,}
\end{equation}
where we took into account that $p^0 = 0$, and where the powers of $(p-q)^2$ and $q^2$ arise in the Fourier transform of the graviton propagators~\eqref{eq_propagators}, and are thus limited to $a,b \in \{1,2\}$. The integral over $q^0$ is performed using the Cauchy residue theorem, closing the integration contour in the upper or lower half plane depending on the sign of $\tau$. Namely, since the only other dependence of the integrand on $q^0$ has the form $(q^0)^c \exp\left( - \mathi q^0 \tau \right)$ for some integer $c$, for $\tau > 0$ the contribution of the arc at infinity in the lower half plane vanishes while for $\tau < 0$ the contribution of the arc at infinity in the upper half plane vanishes. We can combine the result of both branches, and the $q^0$ integral results in
\begin{splitequation}
\label{eq_q0int_residue}
&\int \frac{(q^0)^c \exp\left( - \mathi q^0 \tau \right)}{\left[ \vec{q}^2 - (q^0)^2 + \left[ \left( \vec{p} - \vec{q} \right)^2 - \vec{q}^2 \right] \xi - \mathi \epsilon \right]^{a+b}} \total q^0 \\
&\quad= 2 \pi \mathi \frac{(-1)^{a+b} \left[ \Theta(\tau) + (-1)^c \Theta(-\tau) \right]}{(a+b-1)!} \lim_{q^0 \to \sqrt{ \Xi }} \frac{\total^{a+b-1}}{\total (q^0)^{a+b-1}} \left[ \frac{(q^0)^c \exp\left( - \mathi q^0 \abs{\tau} \right)}{\left( q^0 + \sqrt{ \Xi } \right)^{a+b}} \right] \eqend{,}
\end{splitequation}
where
\begin{equation}
\Xi \equiv \vec{q}^2 + \left[ \left( \vec{p} - \vec{q} \right)^2 - \vec{q}^2 \right] \xi - \mathi \epsilon \eqend{.}
\end{equation}
The subsequent $\tau$ integral is convergent because of the $\mathi \epsilon$ prescription, and we can actually perform it before taking the derivatives in~\eqref{eq_q0int_residue}, using
\begin{equation}
\lim_{\epsilon \to 0} \int \Theta(\pm \tau) \exp\left[ - \mathi ( q^0 - \mathi \epsilon ) \, \abs{\tau} \right] \abs{\tau} \total \tau = - \frac{1}{(q^0)^2} \eqend{.}
\end{equation}
Taking all together, it follows that
\begin{splitequation}
&\iint \frac{(q^0)^c \exp\left( - \mathi q^0 \tau \right)}{[ (p-q)^2 - \mathi \epsilon ]^a ( q^2 - \mathi \epsilon )^b} \total q^0 \abs{\tau} \total \tau \\
&\quad= - 2 \pi \mathi \frac{(-1)^{a+b} \left[ 1 + (-1)^c \right]}{\Gamma(a) \Gamma(b)} \int_0^1 \xi^{a-1} (1-\xi)^{b-1} \lim_{q^0 \to \sqrt{ \Xi }} \frac{\total^{a+b-1}}{\total (q^0)^{a+b-1}} \left[ \frac{(q^0)^{c-2}}{\left( q^0 + \sqrt{ \Xi } \right)^{a+b}} \right] \total \xi \\
&\quad= \pi \mathi \frac{\left[ 1 + (-1)^c \right]}{\Gamma(a) \Gamma(b)} \frac{\Gamma\left( a + b - \frac{c-1}{2} \right)}{\Gamma\left( \frac{3-c}{2} \right)} \int_0^1 \xi^{a-1} (1-\xi)^{b-1} \Xi^{\frac{c-1}{2} - a - b} \total \xi \eqend{.} \raisetag{2.4em}
\end{splitequation}
Finally, we have to perform the integrals over the spatial components $\vec{q}$ and $\vec{p}$, and subsequently over the Feynman parameter $\xi$. Since all the $\vec{q}$ dependence is contained in $\Xi$, we first shift $\vec{q} \to \vec{q} + \xi \vec{p}$, which leads to
\begin{equation}
\Xi \to \vec{q}^2 + \xi (1-\xi) \vec{p}^2 \eqend{.}
\end{equation}
The integral over $\vec{q}$ can then be done with the help of another well-known formula
\begin{equation}
\int \left( \vec{q}^2 + m^2 \right)^a \left( \vec{q}^2 \right)^b \frac{\total^{n-1} \vec{q}}{(2 \pi)^{n-1}} = \frac{1}{(4 \pi)^\frac{n-1}{2}} \frac{\Gamma\left( -a-b-\frac{n-1}{2} \right) \Gamma\left( b + \frac{n-1}{2} \right)}{\Gamma(-a) \Gamma\left( \frac{n-1}{2} \right)} (m^2)^{a+b+\frac{n-1}{2}}
\end{equation}
(see for example~\cite{smirnov2004}, where a Wick rotation is again unnecessary), and the remaining integral over $\vec{p}$ can be done with the previous formula~\eqref{eq_p2_fouriertrafo}, again converting factors of $\vec{p}_\mu$ into derivatives with respect to $\vec{x}^\mu$. In the end, a rational function of $\xi$ is left, which is also easily integrated. The final result for the contribution from the fluctuating worldline to the expectation value of our invariant observable $\expect{ \mathcal{H}_{\mu\nu}(x) }$ is then given by
\begin{splitequation}
\label{eq_corr_worldline_pre}
&- \frac{20 G_\text{N}^2 m}{3 \pi r^3} \left( 3 \frac{\vec{x}_\mu \vec{x}_\nu}{r^2} - \bar\eta_{\mu\nu} \right) \left( - \frac{1}{n-4} + \gamma + \ln(8 \pi) + \ln \mu^2 \right) \\
&\hspace{0.7em}- \frac{20 G_\text{N}^2 m}{3 \pi r^3} \left( 3 \frac{\vec{x}_\mu \vec{x}_\nu}{r^2} - \bar\eta_{\mu\nu} \right) \ln \left( \frac{r^2}{\mu^2} \right) - \frac{43 G_\text{N}^2 m}{6 \pi r^3} v_\mu v_\nu + \frac{727 G_\text{N}^2 m}{12 \pi r^3} \frac{\vec{x}_\mu \vec{x}_\nu}{r^2} - \frac{65 G_\text{N}^2 m}{4 \pi r^3} \bar\eta_{\mu\nu} \eqend{.}
\end{splitequation}
As for the graviton and ghost loop corrections, we renormalize by adding the homogeneous solution $c \, \vec{x}^\mu \, r^{1-n}$ to the $X^{(\mu)}$, now with
\begin{equation}
c = \frac{10}{3} \frac{G_\text{N}^2 m}{\pi} \left( - \frac{1}{n-4} + \gamma + \ln(8 \pi) + \ln \mu - \frac{647}{240} \right) + \bigo{n-4} \eqend{.}
\end{equation}
Adding the resulting contribution $- 2 c \, \eta_{\rho(\mu} \partial_{\nu)} \left( \vec{x}^\rho \, r^{1-n} \right)$ to the result~\eqref{eq_corr_worldline_pre}, the divergent part again cancels and we are left with the finite result
\begin{splitequation}
\label{eq_corr_worldline}
- \frac{10 G_\text{N}^2 m}{3 \pi r^3} \left( 3 \frac{\vec{x}_\mu \vec{x}_\nu}{r^2} - \bar\eta_{\mu\nu} \right) \ln \left( \frac{r^2}{\mu^2} \right) - \frac{43 G_\text{N}^2 m}{6 \pi r^3} v_\mu v_\nu + \frac{31 G_\text{N}^2 m}{18 \pi r^3} \bar\eta_{\mu\nu} \eqend{.}
\end{splitequation}

\subsection{Classical corrections}
\label{sec_class}

\begin{figure}[ht]
\begin{center}
\begin{fmffile}{class_corrections}
	\begin{fmfgraph*}(.4,.15)
		\fmfleft{hin1}
		\fmfright{hout1}
		\fmfv{decoration.shape=circle,decoration.filled=30,decor.size=3.5mm}{hout1}
		\fmf{wiggly}{hin1,hout1}
		\fmfv{label= \small{$ m\kappa$},label.dist=3mm,label.angle=90}{hout1}
	\end{fmfgraph*}
	\hspace{.1\linewidth}
    \begin{fmfgraph*}(.4,.15)
		\fmfdotn{v}{1}
		\fmfleft{hin1}
		\fmfv{decoration.shape=circle,decoration.filled=30,decor.size=3.5mm}{hout1}
		\fmfright{hout1,hout2}
		\fmfv{decoration.shape=circle,decoration.filled=30,decor.size=3.5mm}{hout2}
		\fmf{wiggly}{v1,hout2}
		\fmf{wiggly}{v1,hout1}
		\fmf{wiggly}{hin1,v1}
		\fmfv{label= \small{$ m\kappa$},label.dist=3mm}{hout1,hout2}
		\fmfv{label= \small{$ \kappa$},label.angle=90}{hin1}
		\fmfv{label= \small{$ \kappa$},label.angle=135}{v1}
	\end{fmfgraph*}
	\vspace{.02\linewidth}
\end{fmffile}
\caption{The classical contributions to $\protect\expect{ \mathcal{H}_{\mu\nu}(x) }$ up to order $m^2 \kappa^4 \sim m^2 G_\text{N}^2$. They agree with the expansion of the Schwarzschild metric to that order, with the first one giving the well-known classical Newtonian gravitational potential.}
\label{fig_class_correction}
\end{center}
\end{figure}
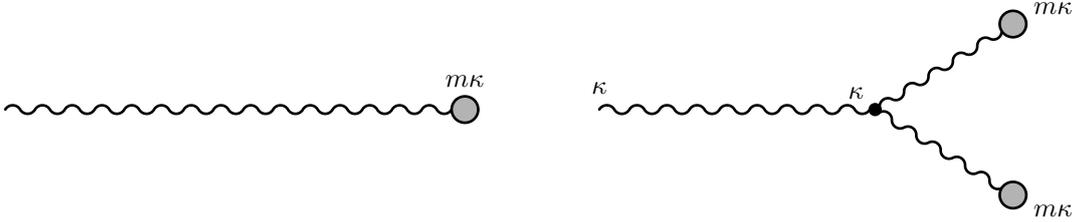

To order $\kappa^4$, there are also perturbative corrections which are purely tree-level, and which agree with the expansion of the Schwarzschild metric for a point source to that order~\cite{duff1973}. The corresponding Feynman diagrams are shown in Fig.~\ref{fig_class_correction}, and are straightforward to compute using the three-graviton vertex that can be read off from~\eqref{eq_sint} and the coupling~\eqref{eq_pp_hmnunu_tmunu_vertex} of the graviton to the point particle stress tensor~\eqref{eq_pp_stresstensor}.

As required for tree-level graphs, the result is finite in $n = 4$ dimensions and does not need renormalisation, and its contribution to the expectation value of our invariant observable $\expect{ \mathcal{H}_{\mu\nu}(x) }$ reads
\begin{equation}
\label{eq_corr_class}
\frac{m G_\text{N}}{r} \left( 2 \bar{\eta}_{\mu\nu} + 2 v_\mu v_\nu \right) + \frac{m^2 G_\text{N}^2}{r^2} \left( - 7 \frac{\vec{x}_\mu \vec{x}_\nu}{r^2} + 5 \bar{\eta}_{\mu\nu} - 2 v_\mu v_\nu \right) \eqend{.}
\end{equation}
Note that this does not yet correspond to the expansion of the Schwarzschild metric in harmonic coordinates since also the corrections from the field-dependent coordinates need to be taken into account, which we do in the next subsection. The exact comparison with the Schwarzschild metric is then given in subsection~\ref{sec_result_class}.

\subsection{Corrections from the field-dependent coordinates}
\label{sec_coord}

\begin{figure}[ht]
\begin{center}
\begin{fmffile}{coord_corrections}
    \begin{subfigure}{0.3\textwidth}
	\begin{fmfgraph*}(.25,.1)
		\fmfdotn{v}{1}
		\fmfleft{hin1}
		\fmfv{decoration.shape=circle,decoration.filled=30,decor.size=3.5mm}{hout1}
		\fmfright{hout1,hout2}
		\fmfv{decoration.shape=circle,decoration.filled=30,decor.size=3.5mm}{hout2}
		\fmf{wiggly}{v1,hout2}
		\fmf{wiggly}{v1,hout1}
		\fmf{dbl_wiggly}{hin1,v1}
		\fmfv{label= \small{$ m\kappa$},label.dist=3mm}{hout1}
		\fmfv{label= \small{$ m\kappa$},label.dist=3mm}{hout2}
		\fmfv{label= \small{$ \kappa^2$},label.angle=90}{hin1}
	\end{fmfgraph*}
	\caption{}
    \label{fig_coord_corrections_a}
	\end{subfigure}
	\hspace{0.03\textwidth}
    \begin{subfigure}{0.3\textwidth}
	\begin{fmfgraph*}(.25,.1)
        \fmfdotn{v}{1}
        \fmfleft{hin1}
        \fmfright{hout1,hout2}
        \fmf{dbl_wiggly}{hin1,v1}
        \fmf{wiggly}{v1,hout2}
        \fmf{wiggly}{v1,hout1}
        \fmf{plain}{hout2,hout1}
        \fmfv{decoration.shape=circle,decoration.filled=full,decoration.size=1.5mm}{hout1,hout2}
        \fmfforce{(.4w,.5h)}{v1}
        \fmfv{label= \small{$ m^{1/2} \kappa$}}{hout2}
        \fmfv{label= \small{$ m^{1/2} \kappa$}}{hout1}
        \fmfv{label= \small{$ \kappa^2$},label.angle=90}{hin1}
    \end{fmfgraph*}
	\caption{}
    \label{fig_coord_corrections_b}
	\end{subfigure}
	\hspace{0.03\textwidth}
    \begin{subfigure}{0.3\textwidth}
    \begin{fmfgraph*}(.25,.1)
            \fmfdotn{v}{2}
            \fmfv{decoration.shape=circle,decoration.filled=30,decor.size=3.5mm}{hout1}
            \fmfleft{hin1}
            \fmfright{hout1}
            \fmf{dbl_wiggly}{hin1,v1}
            \fmf{wiggly,left}{v1,v2}
            \fmf{wiggly,right}{v1,v2}
            \fmf{wiggly}{v2,hout1}
            \fmfforce{(.3w,.5h)}{v1}
            \fmfforce{(.7w,.5h)}{v2}
            \fmfv{label= \small{$ \kappa^2$},label.angle=90}{hin1}
            \fmfv{label= \small{$ \kappa$},label.angle=45}{v2}
            \fmfv{label= \small{$ m\kappa$},label.angle=90,label.dist=3mm}{hout1}
    \end{fmfgraph*}
	\caption{}
    \label{fig_coord_corrections_c}
	\end{subfigure}
	\vspace{.02\linewidth}
\end{fmffile}
\caption{Contributions to $\protect\expect{ \mathcal{H}_{\mu\nu}(x) }$ arising from the field-dependent coordinates. The explicit form of $X_{(2)}$ in \eqref{eqn_field-dep_coords_expression} gives rise to the $X_{(2)}$--graviton--graviton interaction vertex, with the source term $J_{(2)}$ denoted by a double wiggly line since it contains two gravitons. The possible combinations with the other interaction vertices result in the depicted diagrams.}
\label{fig_coord_corrections}
\end{center}
\end{figure}
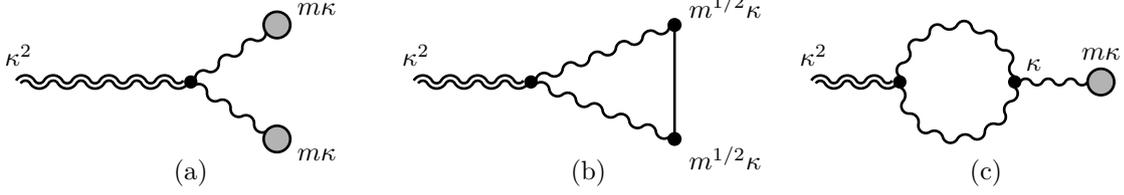

From the definition of the gauge-invariant graviton observable~\eqref{eqn_gauge_inv_H_explicit}, we have further corrections stemming from the field-dependent coordinates. In particular, while $X^\mu_{(1)}$ vanishes by our choice of gauge, we still have corrections coming from the second-order correction $X^\mu_{(2)}$, which are shown in Fig.~\ref{fig_coord_corrections}.

While the diagrams in Fig.~\ref{fig_coord_corrections} almost look like the ones in the previous subsections, there is a small but crucial difference, which is the $X_{(2)}$--graviton--graviton interaction vertex. This vertex does not originate from a term in the interaction $S_\text{int}$, but rather from the definition of the invariant observable $\mathcal{H}_{\mu\nu}(x)$~\eqref{eqn_gauge_inv_H_explicit} through the explicit form of $X^\mu_{(2)}$, involving a source term $J^\mu_{(2)}$ with two gravitons. However, this difference does only affect the tensor algebra, and otherwise the computational strategy for those diagrams is the same as in the previous subsections. The resulting contributions to the expectation value of our invariant observable $\expect{ \mathcal{H}_{\mu\nu}(x) }$ read
\begin{equation}
\label{eq_corr_coord_class}
\frac{m^2 G_\text{N}^2}{r^2} \left( 8 \frac{\vec{x}_\mu \vec{x}_\nu}{r^2} - 4 \bar{\eta}_{\mu\nu} \right)
\end{equation}
for the diagram~\ref{fig_coord_corrections_a},
\begin{splitequation}
\label{eq_corr_coord_worldline_pre}
&\frac{44 G_\text{N}^2 m}{\pi r^3} \left( 3 \frac{\vec{x}_\mu \vec{x}_\nu}{r^2} - \bar\eta_{\mu\nu} \right) \left( - \frac{1}{n-4} + \gamma + \ln(8 \pi) + \ln \mu^2 \right) \\
&\quad+ \frac{44 G_\text{N}^2 m}{\pi r^3} \left( 3 \frac{\vec{x}_\mu \vec{x}_\nu}{r^2} - \bar\eta_{\mu\nu} \right) \ln \left( \frac{r^2}{\mu^2} \right) - \frac{785 G_\text{N}^2 m}{2 \pi r^3} \frac{\vec{x}_\mu \vec{x}_\nu}{r^2} + \frac{203 G_\text{N}^2 m}{2 \pi r^3} \bar\eta_{\mu\nu}
\end{splitequation}
for the diagram~\ref{fig_coord_corrections_b} and
\begin{splitequation}
\label{eq_corr_coord_loop_pre}
&\frac{4 G_\text{N}^2 m}{3 \pi r^3} \left( 3 \frac{\vec{x}_\mu \vec{x}_\nu}{r^2} - \bar\eta_{\mu\nu} \right) \left( - \frac{1}{n-4} + \gamma + \ln(4 \pi) + \ln \mu^2 \right) \\
&\quad+ \frac{4 G_\text{N}^2 m}{3 \pi r^3} \left( 3 \frac{\vec{x}_\mu \vec{x}_\nu}{r^2} - \bar\eta_{\mu\nu} \right) \ln \left( \frac{r^2}{\mu^2} \right) - \frac{34 G_\text{N}^2 m}{3 \pi r^3} \frac{\vec{x}_\mu \vec{x}_\nu}{r^2} + \frac{26 G_\text{N}^2 m}{9 \pi r^3} \bar\eta_{\mu\nu}
\end{splitequation}
for the last diagram~\ref{fig_coord_corrections_c}. Renormalization is again accomplished by adding the term $c \, \vec{x}^\mu \, r^{1-n}$ to the $X^{(\mu)}$, now with
\begin{equation}
c = - 22 \frac{G_\text{N}^2 m}{\pi} \left( - \frac{1}{n-4} + \gamma + \ln(8 \pi) + \ln \mu - \frac{697}{264} \right) + \bigo{n-4} \eqend{.}
\end{equation}
for the result~\eqref{eq_corr_coord_worldline_pre} and
\begin{equation}
c = - \frac{2}{3} \frac{G_\text{N}^2 m}{\pi} \left( - \frac{1}{n-4} + \gamma + \ln(4 \pi) + \ln \mu - \frac{5}{2} \right) + \bigo{n-4} \eqend{.}
\end{equation}
for the result~\eqref{eq_corr_coord_loop_pre}, while~\eqref{eq_corr_coord_class} is finite in $n = 4$ and does not need renormalization. Adding the resulting contribution $- 2 c \, \eta_{\rho(\mu} \partial_{\nu)} \left( \vec{x}^\rho \, r^{1-n} \right)$ to~\eqref{eq_corr_coord_worldline_pre} and~\eqref{eq_corr_coord_loop_pre}, we obtain the finite results
\begin{equation}
\label{eq_corr_coord_worldline}
\frac{22 G_\text{N}^2 m}{\pi r^3} \left( 3 \frac{\vec{x}_\mu \vec{x}_\nu}{r^2} - \bar\eta_{\mu\nu} \right) \ln \left( \frac{r^2}{\mu^2} \right) - \frac{44 G_\text{N}^2 m}{3 \pi r^3} \bar\eta_{\mu\nu}
\end{equation}
for the diagram~\ref{fig_coord_corrections_b} and
\begin{equation}
\label{eq_corr_coord_loop}
\frac{2 G_\text{N}^2 m}{3 \pi r^3} \left( 3 \frac{\vec{x}_\mu \vec{x}_\nu}{r^2} - \bar\eta_{\mu\nu} \right) \ln \left( \frac{r^2}{\mu^2} \right) - \frac{4 G_\text{N}^2 m}{9 \pi r^3} \bar\eta_{\mu\nu}
\end{equation}
for the last diagram~\ref{fig_coord_corrections_c}.

\subsection{Vanishing corrections}
\label{sec_vanish}

\begin{figure}[ht]
\begin{center}
\begin{fmffile}{zero_contribution_diagrams}
    \begin{subfigure}{0.2\textwidth}
    \begin{fmfgraph*}(.2,.1)
        \fmfdotn{v}{1}
        \fmfleft{hin1}
        \fmfv{decoration.shape=circle,decoration.filled=30,decor.size=3.5mm}{hout1}
        \fmfright{hout1}
        \fmf{wiggly}{hin1,v1,v1,hout1}
        \fmfv{label=\small{ $ m \kappa$},label.dist=3mm,label.angle=90}{hout1}
        \fmfv{label= \small{$ \kappa$},label.angle=90}{hin1}
        \fmfv{label= \small{$ \kappa^2$},label.angle=-90}{v1}
    \end{fmfgraph*}
	\caption{}
    \label{fig_zero_contribution_diagrams_a}
    \end{subfigure}
    \hspace{.05\linewidth}
    \begin{subfigure}{0.2\textwidth}
    \begin{fmfgraph*}(.2,.1)
        \fmfdotn{v}{1}
        \fmfleft{hin1}
        \fmfright{hout1}
        \fmfv{decoration.shape=circle,decoration.filled=30,decor.size=3.5mm}{hout1}
        \fmf{dbl_wiggly}{hin1,v1}
        \fmf{wiggly}{v1,v1,hout1}
        \fmfv{label=\small{ $ m \kappa$},label.angle=90,label.dist=3mm}{hout1}
        \fmfv{label= \small{$ \kappa^3$},label.angle=90}{hin1}
    \end{fmfgraph*}
	\caption{}
    \label{fig_zero_contribution_diagrams_b}
    \end{subfigure}
    \hspace{.05\linewidth}
    \begin{subfigure}{0.2\textwidth}
	\begin{fmfgraph*}(.2,.1)
		\fmfdotn{v}{2}
		\fmfleft{hin1}
		\fmfright{hout1}
		\fmfv{decoration.shape=circle,decoration.filled=30,decor.size=3.5mm}{hout1}
		\fmf{wiggly}{hin1,v1}
		\fmf{wiggly}{v2,hout1}
		\fmf{plain}{v1,v2}
		\fmfv{label= \small{$ m^{1/2} \kappa$},label.angle=90}{v1}
		\fmfv{label= \small{$ m^{1/2} \kappa$},label.angle=-90}{v2}
		\fmfv{label=\small{ $ m \kappa$}, label.angle=90,label.dist=3mm}{hout1}
		\fmfv{label= \small{$ \kappa$}, label.angle=90}{hin1}
	\end{fmfgraph*}
	\caption{}
    \label{fig_zero_contribution_diagrams_c}
    \end{subfigure}
    \hspace{.05\linewidth}
    \begin{subfigure}{0.2\textwidth}
	\begin{fmfgraph*}(.2,.1)
        \fmfdotn{v}{2}
        \fmfleft{hin1}
        \fmfright{hout1}
        \fmfv{decoration.shape=circle,decoration.filled=30,decor.size=3.5mm}{hout1}
        \fmf{wiggly}{hin1,v1}
        \fmf{wiggly}{v2,hout1}
        \fmf{plain,right}{v1,v2}
        \fmf{plain,left}{v1,v2}
        \fmfforce{(.3w,.5h)}{v1}
        \fmfforce{(.7w,.5h)}{v2}
        \fmfv{label= \small{$ \kappa$}, label.angle=45}{v2}
        \fmfv{label= \small{$ \kappa$}, label.angle=135}{v1}
        \fmfv{label=\small{ $ m \kappa$}, label.angle=90,label.dist=3mm}{hout1}
        \fmfv{label= \small{$ \kappa$}, label.angle=90}{hin1}
	\end{fmfgraph*}
	\caption{}
    \label{fig_zero_contribution_diagrams_d}
    \end{subfigure} \\
    \vspace{0.02\linewidth}

    \begin{subfigure}{0.2\textwidth}
	\begin{fmfgraph*}(.2,.15)
		\fmfdotn{v}{2}
		\fmfv{decoration.shape=circle,decoration.filled=30,decor.size=3.5mm}{hout1}
		\fmfleft{hin1}
		\fmfright{hout1}
		\fmftop{top}
		\fmf{dbl_wiggly}{hin1,v1}
		\fmf{wiggly}{v1,hout1}
		\fmf{wiggly}{v2,v1}
		\fmf{wiggly,right}{v2,top}
		\fmf{wiggly,left}{v2,top}
		\fmfforce{(.5w,.1h)}{v1}
		\fmfforce{(0w,.1h)}{hin1}
		\fmfforce{(1w,.1h)}{hout1}
		\fmfforce{(.5w,.5h)}{v2}
		\fmfv{label= \small{$ \kappa^2$},label.angle=90}{hin1}
		\fmfv{label=\small{ $ m \kappa$},label.angle=90,label.dist=3mm}{hout1}
		\fmfv{label= \small{$ \kappa$},label.dist=2mm,label.angle=-40}{v2}
	\end{fmfgraph*}
	\caption{}
    \label{fig_zero_contribution_diagrams_e}
    \end{subfigure}
    \hspace{.05\linewidth}
    \begin{subfigure}{0.2\textwidth}
	\begin{fmfgraph*}(.2,.15)
		\fmfdotn{v}{2}
		\fmfv{decoration.shape=circle,decoration.filled=30,decor.size=3.5mm}{hout1}
		\fmfleft{hin1}
		\fmfright{hout1}
		\fmftop{top}
		\fmf{wiggly}{hin1,v1}
		\fmf{wiggly}{v1,hout1}
		\fmf{wiggly}{v2,v1}
		\fmf{wiggly,plain,right}{v2,top}
		\fmf{wiggly,left}{v2,top}
		\fmfforce{(.5w,.1h)}{v1}
		\fmfforce{(0w,.1h)}{hin1}
		\fmfforce{(1w,.1h)}{hout1}
		\fmfforce{(.5w,.5h)}{v2}
		\fmfv{label= \small{$ \kappa$},label.angle=90}{hin1}
		\fmfv{label=\small{ $ m \kappa$},label.angle=90,label.dist=3mm}{hout1}
		\fmfv{label= \small{$ \kappa$},label.dist=2mm,label.angle=-40}{v2}
		\fmfv{label= \small{$ \kappa$},label.dist=2mm,label.angle=-90}{v1}
	\end{fmfgraph*}
	\caption{}
    \label{fig_zero_contribution_diagrams_f}
    \end{subfigure}
    \hspace{.05\linewidth}
    \begin{subfigure}{0.2\textwidth}
	\begin{fmfgraph*}(.2,.15)
		\fmfdotn{v}{2}
		\fmfv{decoration.shape=circle,decoration.filled=30,decor.size=3.5mm}{hout1}
		\fmfleft{hin1}
		\fmfright{hout1}
		\fmftop{top}
		\fmf{dbl_wiggly}{hin1,v1}
		\fmf{wiggly}{v1,hout1}
		\fmf{wiggly}{v2,v1}
		\fmf{plain,right}{v2,top}
		\fmf{plain,left}{v2,top}
		\fmfforce{(.5w,.1h)}{v1}
		\fmfforce{(0w,.1h)}{hin1}
		\fmfforce{(1w,.1h)}{hout1}
		\fmfforce{(.5w,.5h)}{v2}
		\fmfv{label= \small{$ \kappa^2$},label.angle=90}{hin1}
		\fmfv{label=\small{ $ m \kappa$},label.angle=90,label.dist=3mm}{hout1}
		\fmfv{label= \small{$ \kappa$},label.dist=2mm,label.angle=-40}{v2}
	\end{fmfgraph*}
	\caption{}
    \label{fig_zero_contribution_diagrams_g}
    \end{subfigure}
    \hspace{.05\linewidth}
    \begin{subfigure}{0.2\textwidth}
	\begin{fmfgraph*}(.2,.15)
		\fmfdotn{v}{2}
		\fmfv{decoration.shape=circle,decoration.filled=30,decor.size=3.5mm}{hout1}
		\fmfleft{hin1}
		\fmfright{hout1}
		\fmftop{top}
		\fmf{wiggly}{hin1,v1}
		\fmf{wiggly}{v1,hout1}
		\fmf{wiggly}{v2,v1}
		\fmf{plain,right}{v2,top}
		\fmf{plain,left}{v2,top}
		\fmfforce{(.5w,.1h)}{v1}
		\fmfforce{(0w,.1h)}{hin1}
		\fmfforce{(1w,.1h)}{hout1}
		\fmfforce{(.5w,.5h)}{v2}
		\fmfv{label= \small{$ \kappa$},label.angle=90}{hin1}
		\fmfv{label=\small{ $ m \kappa$},label.angle=90,label.dist=3mm}{hout1}
		\fmfv{label= \small{$ \kappa$},label.dist=2mm,label.angle=-40}{v2}
		\fmfv{label= \small{$ \kappa$},label.dist=2mm,label.angle=-90}{v1}
	\end{fmfgraph*}
	\caption{}
    \label{fig_zero_contribution_diagrams_h}
    \end{subfigure} \\
    \vspace{0.02\linewidth}
    
    \begin{subfigure}{0.2\textwidth}
	\begin{fmfgraph*}(.2,.1)
		\fmfdotn{v}{1}
        \fmfleft{hin1}
        \fmfright{hout1,hout2}
        \fmf{wiggly}{hin1,v1}
        \fmf{plain}{v1,hout2}
        \fmf{plain}{v1,hout1}
        \fmf{wiggly}{hout2,hout1}
        \fmfv{decoration.shape=circle,decoration.filled=full,decoration.size=1.5mm}{hout1,hout2}
        \fmfforce{(.4w,.5h)}{v1}
        \fmfv{label= \small{$ m^{1/2} \kappa$}}{hout2}
        \fmfv{label= \small{$ m^{1/2} \kappa$}}{hout1}
        \fmfv{label= \small{$ \kappa$},label.angle=90}{hin1}
        \fmfv{label= \small{$ \kappa$},label.dist=3mm,label.angle=135}{v1}
    \end{fmfgraph*}
	\caption{}
    \label{fig_zero_contribution_diagrams_i}
    \end{subfigure}
    \hspace{.05\linewidth}
    \begin{subfigure}{0.2\textwidth}
    \begin{fmfgraph*}(.2,.1)
        \fmfdotn{v}{2}
        \fmfleft{hin1}
        \fmfright{hout1}
        \fmf{wiggly}{hin1,v1}
        \fmf{plain}{v1,v2}
        \fmf{plain,right}{v2,hout1}
        \fmf{wiggly,left}{v2,hout1}
        \fmfforce{(.6w,.5h)}{v2}
        \fmfv{decoration.shape=circle,decoration.filled=full,decoration.size=1.5mm}{hout1}
        \fmfv{label= \small{$ m^{1/2} \kappa$},label.dist=2mm,label.angle=-90}{v1}
        \fmfv{label= \small{$ m^{1/2} \kappa$},label.dist=2mm}{hout1}
        \fmfv{label= \small{$ \kappa$},label.angle=-90}{hin1}
        \fmfv{label= \small{$ \kappa$},label.dist=3mm,label.angle=135}{v2}
    \end{fmfgraph*}
	\caption{}
    \label{fig_zero_contribution_diagrams_j}
    \end{subfigure}
    \vspace{0.02\linewidth}
\end{fmffile}
\caption{These contributions vanish due to a massless closed graviton loop (a, b, e, f), closed-$y$ loops (g, h), and graviton propagators that propagate from the worldline to the worldline (c, d). In addition, the diagrams i and j result in scaleless integrals and vanish as well.}
\label{fig_zero_contribution_diagrams}
\end{center}
\end{figure}
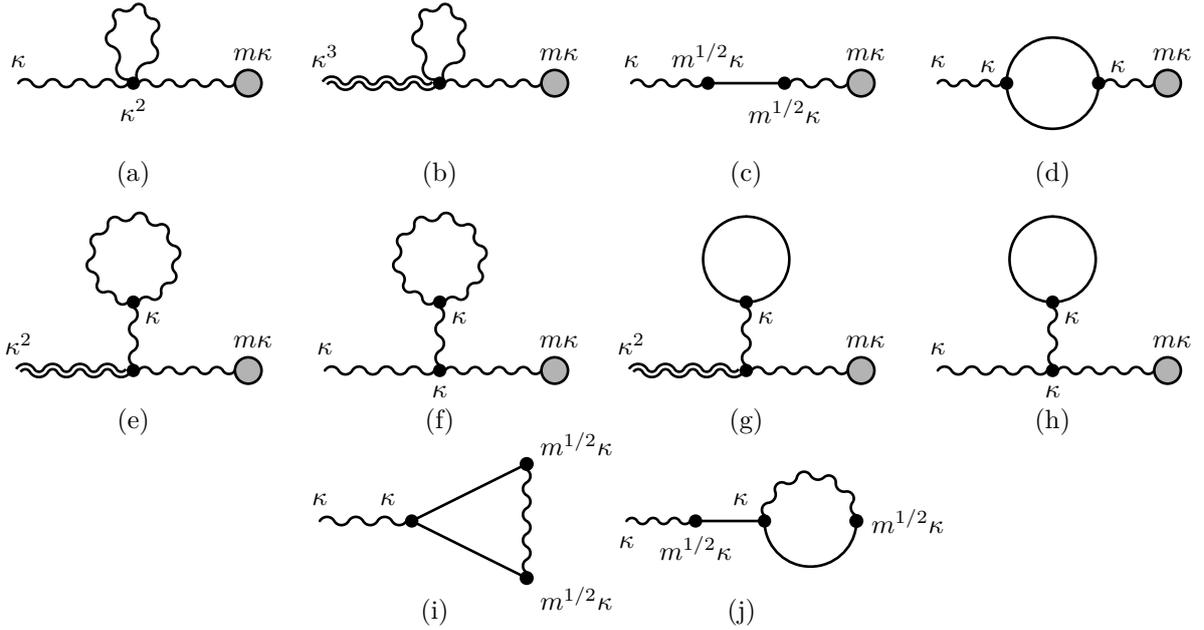

Lastly, we have a look at the (many) terms that in principle arise from the Gell-Mann--Low formula, but turn out to give zero contribution to the result for various reasons. The corresponding Feynman diagrams are depicted in Fig.~\ref{fig_zero_contribution_diagrams}.
Since in dimensional regularization scaleless integrals vanish (are regularized to zero)~\cite{leibbrandt1975}, any $n$-dimensional closed loop in which massless particles such as the graviton propagate does vanish, which concerns the diagrams~\ref{fig_zero_contribution_diagrams_a}, \subref{fig_zero_contribution_diagrams_b}, \subref{fig_zero_contribution_diagrams_e} and~\subref{fig_zero_contribution_diagrams_f}. However, closed world-line-perturbation loops do not vanish by this argument, since the integration is over the time parameter $\tau$ only which has the fixed dimension $1$. Nevertheless, since the $y$ propagator~\eqref{eq_propagators} only depends on the difference in time parameters, any closed $y$ loop (without derivatives acting on the propagator) vanishes as well, which concerns the diagrams~\ref{fig_zero_contribution_diagrams_g} and~\subref{fig_zero_contribution_diagrams_h}. On the other hand, the two diagrams~\ref{fig_zero_contribution_diagrams_i} and~\subref{fig_zero_contribution_diagrams_j} contain a graviton propagator sandwiched between interaction vertices which otherwise only contain worldline perturbations. Following the same strategy as in subsection~\ref{sec_worldline} to ccompute the loop, we obtain scaleless $(n-1)$-dimensional integrals over the graviton propagator momentum, which then again vanish in dimensional regularization. Finally, the diagrams~\ref{fig_zero_contribution_diagrams_c} and~\subref{fig_zero_contribution_diagrams_d} vanish because they contain a graviton propagating from the wordline to itself, connecting the worldline perturbation with the point particle's stress tensor. As in subsection~\ref{sec_loop}, the integral over the time $\tau$ of the point particle's stress tensor sets the time component of the graviton propagator momentum $p^0$ to zero, and since the worldline perturbation is also purely time-dependent, we have to integrate the graviton propagator over its spatial momentum $\vec{p}$. However, this is again a scaleless integral which vanishes in dimensional regularization.

\section{Results}
\label{sec_results}

Adding up the contributions~\eqref{eq_corr_class} and~\eqref{eq_corr_coord_class}, we obtain the classical corrections to the Newtonian potential up to order $G_\text{N}^2$, which we analyse in subsection~\ref{sec_result_class}. The other contributions are all quantum and proportional to $\hbar$, and fall naturally into two classes: the contributions~\eqref{eq_corr_loop} and~\eqref{eq_corr_coord_loop} stemming from graviton and ghost loops, and which would also be present if instead of a point particle we would have coupled a scalar field to gravity, as in the previous works~\cite{kirilinkhriplovich2002,khriplovichkirilin2003,bjerrumbohrdonoghueholstein2003a,bjerrumbohrdonoghueholstein2003b}. The second type of contributions~\eqref{eq_corr_worldline} and~\eqref{eq_corr_coord_worldline} on the other hand arise from the fluctuating worldline of the point particle, and are unique to the present situation. We analyse both kinds of quantum corrections in subsection~\ref{sec_result_quantum}.

\subsection{Classical contributions}
\label{sec_result_class}

The classical contributions~\eqref{eq_corr_class} and~\eqref{eq_corr_coord_class} to the expectation value of our invariant graviton observable $\expect{ \mathcal{H}_{\mu\nu}(x) }$ sum up to
\begin{equation}
\label{eq_result_class}
\expect{ \mathcal{H}_{\mu\nu}(x) }^\text{class} = \frac{G_\text{N} m}{r} \left( 2 \bar{\eta}_{\mu\nu} + 2 v_\mu v_\nu \right) + \frac{G_\text{N}^2 m^2}{r^2} \left( \frac{\vec{x}_\mu \vec{x}_\nu}{r^2} + \bar{\eta}_{\mu\nu} - 2 v_\mu v_\nu \right) \eqend{,}
\end{equation}
where we recall that both $\vec{x}_\mu$ and $\bar{\eta}_{\mu\nu}$~\eqref{eq_bareta_def} are purely spatial. Since they give the classical gravitational response to a point source, they should agree with the expansion to order $G^2$ of the corresponding exact solution of Einstein gravity~\cite{duff1973}, the well-known Schwarzschild metric. Because we have chosen a coordinate system corresponding to harmonic coordinates~\eqref{eqn_gen_harm_coords_ansatz}, we also need to take the Schwarzschild metric in harmonic coordinates. This is obtained from the usual Schwarzschild coordinates by a simple shift $r \to r + G_\text{N} m$~\cite{duff1973} and reads
\begin{equation}
\total s^2 = g_{\mu\nu} \total x^\mu \total x^\nu = - \frac{r - G_\text{N} m}{r + G_\text{N} m} \total t^2 + \left( 1 + \frac{G_\text{N} m}{r} \right)^2 \total \vec{x}^2 + \frac{r + G_\text{N} m}{r - G_\text{N} m} \frac{G_\text{N}^2 m^2}{r^4} \left( \vec{x} \cdot \total \vec{x} \right)^2 \eqend{,}
\end{equation}
where as usual $r^2 \equiv \vec{x}^2$. Expanding this to second order in $G_\text{N}$, we obtain
\begin{equation}
g_{\mu\nu} = \eta_{\mu\nu} + \left( \frac{2 G_\text{N} m}{r} - \frac{2 G_\text{N}^2 m^2}{r^2} \right) v_\mu v_\nu + \left( \frac{2 G_\text{N} m}{r} + \frac{G_\text{N}^2 m^2}{r^2} \right) \bar{\eta}_{\mu\nu} + \frac{G_\text{N}^2 m^2}{r^4} \vec{x}_\mu \vec{x}_\nu \eqend{,}
\end{equation}
which agrees with~\eqref{eq_result_class} after subtracting the background metric $\eta_{\mu\nu}$. In particular, the classical contribution to the corrected Newtonian potential reads
\begin{equation}
V^\text{class}(r) \equiv - \frac{1}{2} \expect{ \mathcal{H}_{00}(x) }^\text{class} = - \frac{G_\text{N} m}{r} \left( 1 - \frac{G_\text{N} m}{r} \right) \eqend{.}
\end{equation}
The classical contributions of higher order in $G_\text{N}$ are called post-Newtonian, and feature prominently in the computation of black hole inspirals~\cite{levi2018}. However, since we are mainly interested in quantum corrections, here they serve merely as a check on the computation.

\subsection{Quantum contributions}
\label{sec_result_quantum}

The graviton and ghost loop corrections~\eqref{eq_corr_loop} including the corrections from the field-dependent coordinates~\eqref{eq_corr_coord_loop} to the expectation value of our invariant graviton observable $\expect{ \mathcal{H}_{\mu\nu}(x) }$ sum up to
\begin{equation}
\label{eq_result_quant_loop}
\expect{ \mathcal{H}_{\mu\nu}(x) }^\text{loop} = \frac{92 G_\text{N}^2 m}{15 \pi r^3} \left( 3 \frac{\vec{x}_\mu \vec{x}_\nu}{r^2} - \bar\eta_{\mu\nu} \right) \ln \left( \frac{r^2}{\mu^2} \right) + \frac{41 G_\text{N}^2 m}{5 \pi r^3} v_\mu v_\nu - \frac{307 G_\text{N}^2 m}{45 \pi r^3} \bar\eta_{\mu\nu} \eqend{,}
\end{equation}
which gives the contribution
\begin{equation}
\label{eq_result_vloop}
V^\text{loop}(r) \equiv - \frac{1}{2} \expect{ \mathcal{H}_{00}(x) }^\text{loop} = - \frac{41}{10 \pi} \frac{G_\text{N}^2 m}{r^3} \eqend{.}
\end{equation}
to the corrected Newtonian potential. This correction coincides \emph{exactly} with the result~\eqref{intro_constant_c} obtained using the inverse scattering method~\cite{khriplovichkirilin2003,bjerrumbohrdonoghueholstein2003b}. That is, in complete analogy to the matter loop corrections one can obtain a gauge-invariant result in different ways, which all must agree for consistency. Moreover, this universal part of the result does not depend on how one determines the corrections, whether by coupling gravity to scalar fields (as in the inverse scattering method, which computes $S$-matrix elements for the scattering of scalar fields) or to a point particle (as we do). However, for a point particle there are also corrections~\eqref{eq_corr_worldline} and~\eqref{eq_corr_coord_worldline} stemming from fluctuations of the worldline, which obviously do not arise for a scalar field. Summing those gives the contribution
\begin{equation}
\label{eq_result_quant_worldline}
\expect{ \mathcal{H}_{\mu\nu}(x) }^\text{worldline} = \frac{56 G_\text{N}^2 m}{3 \pi r^3} \left( 3 \frac{\vec{x}_\mu \vec{x}_\nu}{r^2} - \bar\eta_{\mu\nu} \right) \ln \left( \frac{r^2}{\mu^2} \right) - \frac{43 G_\text{N}^2 m}{6 \pi r^3} v_\mu v_\nu - \frac{233 G_\text{N}^2 m}{18 \pi r^3} \bar\eta_{\mu\nu}
\end{equation}
to the expectation value of our invariant observable, and
\begin{equation}
\label{eq_result_vworldline}
V^\text{worldline}(r) \equiv - \frac{1}{2} \expect{ \mathcal{H}_{00}(x) }^\text{worldline} = \frac{43 G_\text{N}^2 m}{12 \pi r^3}
\end{equation}
to the corrected Newtonian potential. If one is only interested in matter fluctations (formalized in a large-$N$ expansion~\cite{tomboulis1977,hartlehorowitz1981,hurouraverdaguer2004}), the worldline corrections are of higher order and do not contribute, but they are of the same order as graviton fluctuations and we must include them here. We interpret this correction as the effect of a quantum fluctuation or ``Zitterbewegung''~\cite{schroedinger1930,thaller,kobakhidzemanningtureanu2016,ecksteinfrancomiller2017} of the point particle reacting to the gravitational interaction, with the caveat that this is to be seen by way of an analogy.

We would like to stress here that while the need for renormalization of loop corrections introduces the (a priori arbitrary) renormalisation scale $\mu$ in the results~\eqref{eq_result_quant_loop} and~\eqref{eq_result_quant_worldline} for the expectation value of the invariant metric perturbation, the quantum corrections to the Newtonian potential~\eqref{eq_result_vloop} and~\eqref{eq_result_vworldline} are independent of this scale. They are therefore completely fixed, and constitute an unambiguous prediction of the low-energy effective quantum field theory of gravity~\cite{donoghue1994b,burgess2004}. The same applies to the spatial trace $\bar\eta^{\mu\nu} \expect{ \mathcal{H}_{\mu\nu}(x) }$, since one easily checks that the trace $- 2 c \, \partial_\mu \left( \vec{x}^\mu \, r^{1-n} \right)$ of the homogeneous solution's contribution to the expectation value vanishes.

\subsection{Discussion and Outlook}
\label{sec_discussion}

\begin{figure}[ht]
\begin{center}
\includegraphics[scale=0.5]{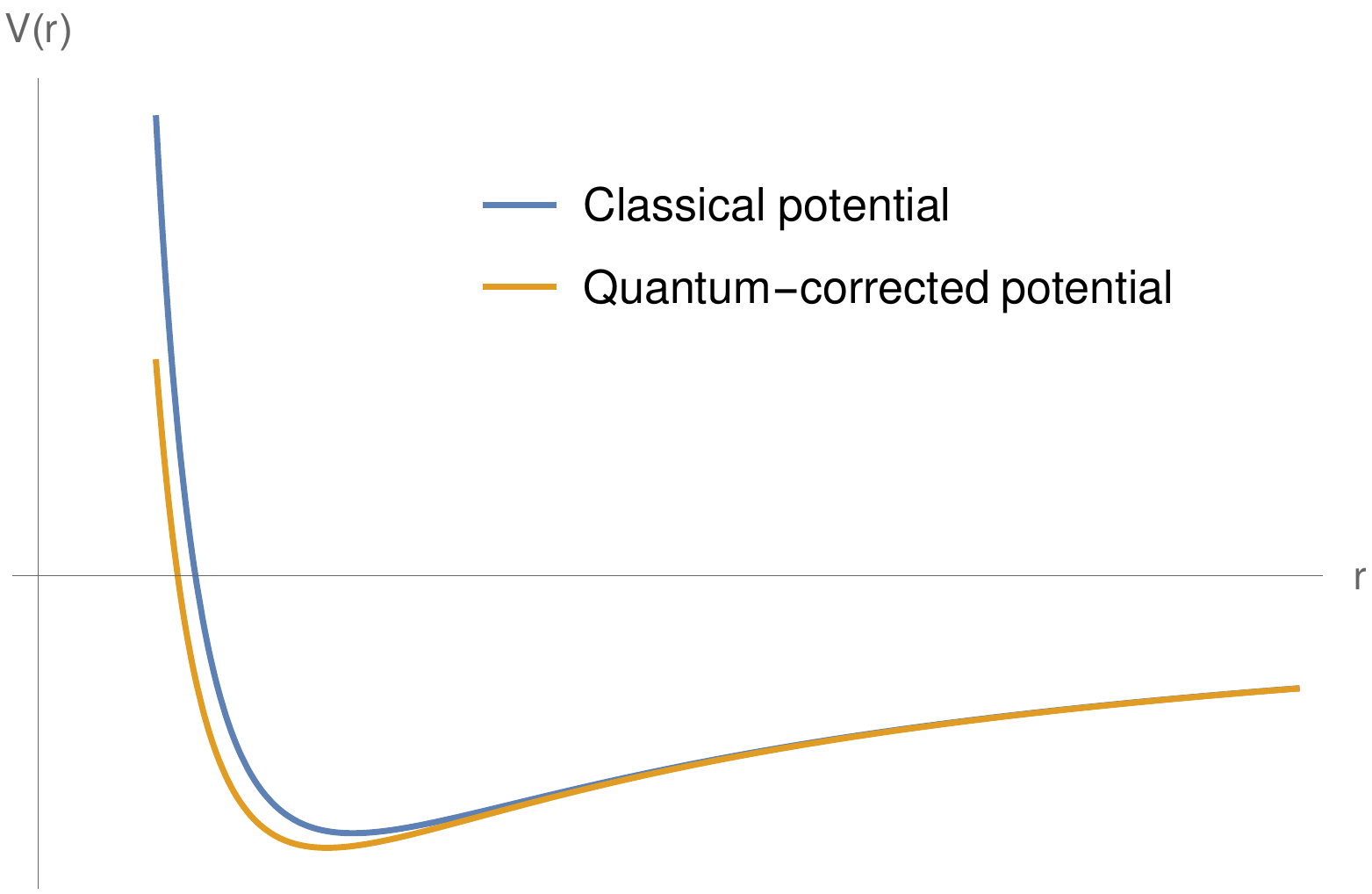}
\caption{Qualitative comparison between the classical (post-)Newtonian gravitational potential and the quantum-corrected one, where the size of the quantum corrections is vastly exaggerated to be visible.}
\label{fig_result_newtonpotential}
\end{center}
\end{figure}

We have computed the corrections to the gravitational potential of a point particle coupled to perturbative quantum gravity. Summing all contributions and restoring factors of $\hbar$, we have to order $G_\text{N}^2$
\begin{equation}
\label{eq_result_newtonpotential}
V(r) = - \frac{G_\text{N} m}{r} \left( 1 - \frac{G_\text{N} m}{r} + \frac{41}{10 \pi} \frac{\hbar G_\text{N}}{r^2} - \frac{43}{12 \pi} \frac{\hbar G_\text{N}}{r^2} \right) \eqend{,}
\end{equation}
where the first term is the classical Newtonian gravitational potential, the second term the first classical post-Newtonian correction, the third term is due to the vacuum polarisation caused by virtual gravitons and ghosts, and the last one stems from the point particle's quantum-mechanical analog of a ``Zitterbewegung'', or more generally, fluctuation of position and energy-momentum, as mentioned before. A sketch of $V(r)$ is given in Fig.~\ref{fig_result_newtonpotential}.
We have obtained these corrections by evaluating the expectation value of a gauge-invariant, non-local and relational observable $\mathcal{H}_{\mu\nu}$ corresponding to the metric perturbation around a flat background. Even though all but the last of these corrections have been computed previously, the method that we have presented here seems to have a number of advantages (and one possible disadvantage):
\begin{itemize}
\item Choosing a suitable gauge, one needs to compute much less diagrams than with the inverse scattering method. In fact, it took 10 years from the first computation~\cite{donoghue1994a} until all required diagrams were computed and included in the result~\cite{bjerrumbohrdonoghueholstein2003b,khriplovichkirilin2003}.
\item There does not seem to be any fundamental ambiguity in the definition of the corrected Newtonian potential (see Ref.~\cite{bjerrumbohrdonoghueholstein2003b} and references therein for different definitions employed with other methods). In particular, the well-known definition $V(x) = - \frac{1}{2} \expect{ h_{00}(x) }$ valid in the classical weak gravity approximation~\cite{mtw}, with $h_{00}$ replaced with our invariant observable $\mathcal{H}_{00}$, gives the right results.
\item One can generalize our method to curved spacetimes, where a scattering matrix may not exist. In particular, it would be very interesting to determine corrections to the gravitational potential in de~Sitter spacetime, where one may obtain corrections that grow with the distance from the source.
\item Even though the result is gauge-independent, it depends functionally on the field-dependent coordinate system. While we have chosen generalized harmonic coordinates, it is possible that using different coordinates the result changes. In the end, the coordinates must be chosen such as to reflect the experimental setup; since the harmonic Cartesian coordinates of the background correspond to straight lines, we interpret the field-dependent generalized harmonic coordinates as coordinates that are ``as straight as possible'' given the unavoidable gravitational interaction. This does seem a reasonable choice in flat space, but for a curved spacetime other coordinates might be more appropriate, for example the co-moving ones of Refs.~\cite{froeblima2018,froeb2019,lima2020} in an inflationary background, or the analogue of the geodesic light-cone coordinates~\cite{gasperinietal2011,froeblima2021}.
\end{itemize}

Let us note that one obtains also corrections to the other gravitational potentials, namely to the spatial trace
\begin{equation}
\expect{ \mathcal{H}_{ii}(x) } = \frac{2 G_\text{N} m}{r} \left[ 1 + \frac{2 G_\text{N} m}{r} - \frac{307 G_\text{N}}{30 \pi r^2} - \frac{233 G_\text{N}}{12 \pi r^2} \right]
\end{equation}
and, if one considers particles with spin, also the gravitoelectric/-magnetic potential $\mathcal{H}_{0i}$. The full result can then be interpreted as giving quantum corrections to the Schwarzschild or Kerr black holes. While it has been speculated that quantum corrections can resolve the central singularity at $r = 0$ (see for example Ref.~\cite{modesto2005}), this is certainly not the case for our one-loop correction~\eqref{eq_result_newtonpotential} since the quantum corrections are actually more singular at the origin. However, the quantum corrections lead to a displacement of the black hole horizon, for which we need to go back from generalized harmonic to Schwarzschild coordinates by reverse shifting $r \to r - G_\text{N} m$. We then have $V(r_\text{H}) = 0$ for
\begin{equation}
r_\text{H} = r_\text{S} - \left( \frac{41}{10 \pi} - \frac{43}{12 \pi} \right) \frac{\hbar}{m}
\end{equation}
with the classical Schwarzschild radius $r_\text{S} = 2 G_\text{N} m$. That is, the quantum corrections lead to a decrease in the horizon radius with respect to the classical theory, and they achieve this by strengthening the gravitational force at small distances, as can also be seen in Fig.~\ref{fig_result_newtonpotential}.

In the future, it would be very interesting to generalize our computation to higher loops and curved backgrounds, and to particles with spin. While the effects of spinning matter on the gravitational potentials have been investigated in both approaches~\cite{bjerrumbohrdonoghueholstein2003a,froeb2016} and the result agrees, the effects of graviton loops have only been determined using the inverse scattering method~\cite{kirilinkhriplovich2002,bjerrumbohrdonoghueholstein2003a,khriplovichkirilin2003} so far. However, since for a spinless particle we have obtained the same result as with the inverse scattering method, plus an additional term coming from the fact that we employ a point particle source, it is highly probable that the same happens for a spinning particle. More interesting results can be obtained for curved backgrounds, for example de~Sitter space which is relevant as a background spacetime in inflationary models. Quantum corrections in this spacetime due to matter loops have been computed~\cite{wangwoodard2015b,parkprokopecwoodard2016,froebverdaguer2016a,froebverdaguer2017}, which in addition to the flat-space result feature an additional term that for massless matter fields grows logarithmically with distance. Since gravitons are massless, one would obtain such a logarithmically growing term also for them. In fact, it is expected on physical grounds that in a de~Sitter background quantum loop corrections involving gravitons give one logarithm per loop~\cite{tsamiswoodard1996}. Such a secular growth signals the breakdown of perturbation theory after a sufficiently long time and/or distance, and one would have to employ some sort of resummation procedure to obtain a non-perturbative result, which then might even be observable.

\begin{acknowledgments}
This work has been funded by the Deutsche Forschungsgemeinschaft (DFG, German Research Foundation) --- project nos. 415803368 and 406116891 within the Research Training Group RTG 2522/1.
\end{acknowledgments}

\addcontentsline{toc}{section}{References}
\bibliography{newton_invobs}

\providecommand{\href}[2]{#2}\begingroup\raggedright\begin{thebibliography}{10}

\bibitem{donoghue1994b}
J.F.~Donoghue, \emph{{General relativity as an effective field theory: The
  leading quantum corrections}},
  \href{https://doi.org/10.1103/PhysRevD.50.3874}{\emph{Phys.~Rev.~D}
  {\bfseries 50} (1994) 3874}
  [\href{https://arxiv.org/abs/gr-qc/9405057}{{\ttfamily gr-qc/9405057}}].

\bibitem{burgess2004}
C.P.~Burgess, \emph{{Quantum gravity in everyday life: General relativity as an
  effective field theory}},
  \href{https://doi.org/10.12942/lrr-2004-5}{\emph{Living~Rev.~Rel.} {\bfseries
  7} (2004) 5} [\href{https://arxiv.org/abs/gr-qc/0311082}{{\ttfamily
  gr-qc/0311082}}].

\bibitem{radkowski1970}
A.F.~Radkowski, \emph{{Some Aspects of the Source Description of Gravitation}},
  \href{https://doi.org/10.1016/0003-4916(70)90021-7}{\emph{Ann.~Phys.}
  {\bfseries 56} (1970) 319}.

\bibitem{schwinger1968}
J.S.~Schwinger, \emph{{Sources and gravitons}},
  \href{https://doi.org/10.1103/PhysRev.173.1264}{\emph{Phys.~Rev.} {\bfseries
  173} (1968) 1264}.

\bibitem{duff1974}
M.J.~Duff, \emph{{Quantum corrections to the Schwarzschild solution}},
  \href{https://doi.org/10.1103/PhysRevD.9.1837}{\emph{Phys.~Rev.~D} {\bfseries
  9} (1974) 1837}.

\bibitem{capperduffhalpern1974}
D.M.~Capper, M.J.~Duff and L.~Halpern, \emph{{Photon corrections to the
  graviton propagator}},
  \href{https://doi.org/10.1103/PhysRevD.10.461}{\emph{Phys.~Rev.~D} {\bfseries
  10} (1974) 461}.

\bibitem{capperduff1974}
D.M.~Capper and M.J.~Duff, \emph{{The one-loop neutrino contribution to the
  graviton propagator}},
  \href{https://doi.org/10.1016/0550-3213(74)90582-3}{\emph{Nucl.~Phys.~B}
  {\bfseries 82} (1974) 147}.

\bibitem{donoghue1994a}
J.F.~Donoghue, \emph{{Leading quantum correction to the Newtonian potential}},
  \href{https://doi.org/10.1103/PhysRevLett.72.2996}{\emph{Phys.~Rev.~Lett.}
  {\bfseries 72} (1994) 2996}
  [\href{https://arxiv.org/abs/gr-qc/9310024}{{\ttfamily gr-qc/9310024}}].

\bibitem{muzinichvokos1995}
I.J.~Muzinich and S.~Vokos, \emph{{Long range forces in quantum gravity}},
  \href{https://doi.org/10.1103/PhysRevD.52.3472}{\emph{Phys.~Rev.~D}
  {\bfseries 52} (1995) 3472}
  [\href{https://arxiv.org/abs/hep-th/9501083}{{\ttfamily hep-th/9501083}}].

\bibitem{hamberliu1995}
H.W.~Hamber and S.~Liu, \emph{{On the quantum corrections to the Newtonian
  potential}},
  \href{https://doi.org/10.1016/0370-2693(95)00790-R}{\emph{Phys.~Lett.~B}
  {\bfseries 357} (1995) 51}
  [\href{https://arxiv.org/abs/hep-th/9505182}{{\ttfamily hep-th/9505182}}].

\bibitem{akhundovbelluccishiekh1997}
A.A.~Akhundov, S.~Bellucci and A.~Shiekh, \emph{{Gravitational interaction to
  one loop in effective quantum gravity}},
  \href{https://doi.org/10.1016/S0370-2693(96)01694-2}{\emph{Phys.~Lett.~B}
  {\bfseries 395} (1997) 16}
  [\href{https://arxiv.org/abs/gr-qc/9611018}{{\ttfamily gr-qc/9611018}}].

\bibitem{dalvitmazzitelli1997}
D.A.R.~Dalvit and F.D.~Mazzitelli, \emph{{Geodesics, gravitons and the gauge
  fixing problem}},
  \href{https://doi.org/10.1103/PhysRevD.56.7779}{\emph{Phys.~Rev.~D}
  {\bfseries 56} (1997) 7779}
  [\href{https://arxiv.org/abs/hep-th/9708102}{{\ttfamily hep-th/9708102}}].

\bibitem{duffliu2000a}
M.J.~Duff and J.T.~Liu, \emph{{Complementarity of the Maldacena and
  Randall-Sundrum pictures}},
  \href{https://doi.org/10.1088/0264-9381/18/16/310}{\emph{Class.~Quant.~Grav.}
  {\bfseries 18} (2001) 3207}
  [\href{https://arxiv.org/abs/hep-th/0003237}{{\ttfamily hep-th/0003237}}].

\bibitem{duffliu2000b}
M.J.~Duff and J.T.~Liu, \emph{{Complementarity of the Maldacena and
  Randall-Sundrum pictures}},
  \href{https://doi.org/10.1103/PhysRevLett.85.2052}{\emph{Phys.~Rev.~Lett.}
  {\bfseries 85} (2000) 2025}
  [\href{https://arxiv.org/abs/hep-th/0003237}{{\ttfamily hep-th/0003237}}].

\bibitem{kirilinkhriplovich2002}
G.G.~Kirilin and I.B.~Khriplovich, \emph{{Quantum power correction to the
  Newton law}},
  \href{https://doi.org/10.1134/1.1537290}{\emph{J.~Exp.~Theor.~Phys.}
  {\bfseries 95} (2002) 981}
  [\href{https://arxiv.org/abs/gr-qc/0207118}{{\ttfamily gr-qc/0207118}}].

\bibitem{khriplovichkirilin2003}
I.B.~Khriplovich and G.G.~Kirilin, \emph{{Quantum long range interactions in
  general relativity}},
  \href{https://doi.org/10.1134/1.1777618}{\emph{J.~Exp.~Theor.~Phys.}
  {\bfseries 98} (2004) 1063}
  [\href{https://arxiv.org/abs/gr-qc/0402018}{{\ttfamily gr-qc/0402018}}].

\bibitem{bjerrumbohrdonoghueholstein2003a}
N.E.J.~Bjerrum-Bohr, J.F.~Donoghue and B.R.~Holstein, \emph{{Quantum
  corrections to the Schwarzschild and Kerr metrics}},
  \href{https://doi.org/10.1103/PhysRevD.68.084005}{\emph{Phys.~Rev.~D}
  {\bfseries 68} (2003) 084005}
  [\href{https://arxiv.org/abs/hep-th/0211071}{{\ttfamily hep-th/0211071}}].

\bibitem{bjerrumbohrdonoghueholstein2003b}
N.E.J.~Bjerrum-Bohr, J.F.~Donoghue and B.R.~Holstein, \emph{{Quantum
  gravitational corrections to the nonrelativistic scattering potential of two
  masses}},
  \href{https://doi.org/10.1103/PhysRevD.67.084033}{\emph{Phys.~Rev.~D}
  {\bfseries 67} (2003) 084033}
  [\href{https://arxiv.org/abs/hep-th/0211072}{{\ttfamily hep-th/0211072}}].

\bibitem{satzmazzitellialvarez2005}
A.~Satz, F.D.~Mazzitelli and E.~Alvarez, \emph{{Vacuum polarization around
  stars: Nonlocal approximation}},
  \href{https://doi.org/10.1103/PhysRevD.71.064001}{\emph{Phys.~Rev.~D}
  {\bfseries 71} (2005) 064001}
  [\href{https://arxiv.org/abs/gr-qc/0411046}{{\ttfamily gr-qc/0411046}}].

\bibitem{holsteinross2008}
B.R.~Holstein and A.~Ross, \emph{{Spin Effects in Long Range Gravitational
  Scattering}},  \href{https://arxiv.org/abs/0802.0716}{{\ttfamily 0802.0716}}.

\bibitem{parkwoodard2010}
S.~Park and R.P.~Woodard, \emph{{Solving the Effective Field Equations for the
  Newtonian Potential}},
  \href{https://doi.org/10.1088/0264-9381/27/24/245008}{\emph{Class.~Quant.~Grav.}
  {\bfseries 27} (2010) 245008}
  [\href{https://arxiv.org/abs/1007.2662}{{\ttfamily 1007.2662}}].

\bibitem{marunovicprokopec2011}
A.~Marunovic and T.~Prokopec, \emph{{Time transients in the quantum corrected
  Newtonian potential induced by a massless nonminimally coupled scalar
  field}},
  \href{https://doi.org/10.1103/PhysRevD.83.104039}{\emph{Phys.~Rev.~D}
  {\bfseries 83} (2011) 104039}
  [\href{https://arxiv.org/abs/1101.5059}{{\ttfamily 1101.5059}}].

\bibitem{marunovicprokopec2012}
A.~Marunovic and T.~Prokopec, \emph{{Antiscreening in perturbative quantum
  gravity and resolving the Newtonian singularity}},
  \href{https://doi.org/10.1103/PhysRevD.87.104027}{\emph{Phys.~Rev.~D}
  {\bfseries 87} (2013) 104027}
  [\href{https://arxiv.org/abs/1209.4779}{{\ttfamily 1209.4779}}].

\bibitem{burnspilaftsis2015}
D.~Burns and A.~Pilaftsis, \emph{{Matter Quantum Corrections to the Graviton
  Self-Energy and the Newtonian Potential}},
  \href{https://doi.org/10.1103/PhysRevD.91.064047}{\emph{Phys.~Rev.~D}
  {\bfseries 91} (2015) 064047}
  [\href{https://arxiv.org/abs/1412.6021}{{\ttfamily 1412.6021}}].

\bibitem{bjerrumbohretal2016}
N.E.J.~Bjerrum-Bohr, J.F.~Donoghue, B.R.~Holstein, L.~Plante and P.~Vanhove,
  \emph{{Light-like Scattering in Quantum Gravity}},
  \href{https://doi.org/10.1007/JHEP11(2016)117}{\emph{JHEP} {\bfseries 11}
  (2016) 117} [\href{https://arxiv.org/abs/1609.07477}{{\ttfamily
  1609.07477}}].

\bibitem{froeb2016}
M.B.~Fr{\"o}b, \emph{{Quantum gravitational corrections for spinning
  particles}}, \href{https://doi.org/10.1007/JHEP10(2016)051}{\emph{JHEP}
  {\bfseries 10} (2016) 051}
  [\href{https://arxiv.org/abs/1607.03129}{{\ttfamily 1607.03129}}].

\bibitem{bjerrumbohrdonoghuevanhove2014}
N.E.J.~Bjerrum-Bohr, J.F.~Donoghue and P.~Vanhove, \emph{{On-shell techniques
  and universal results in quantum gravity}},
  \href{https://doi.org/10.1007/JHEP02(2014)111}{\emph{JHEP} {\bfseries 02}
  (2014) 111} [\href{https://arxiv.org/abs/1309.0804}{{\ttfamily 1309.0804}}].

\bibitem{wangwoodard2015b}
C.L.~Wang and R.P.~Woodard, \emph{{One-loop quantum electrodynamic correction
  to the gravitational potentials on de Sitter spacetime}},
  \href{https://doi.org/10.1103/PhysRevD.92.084008}{\emph{Phys.~Rev.~D}
  {\bfseries 92} (2015) 084008}
  [\href{https://arxiv.org/abs/1508.01564}{{\ttfamily 1508.01564}}].

\bibitem{parkprokopecwoodard2016}
S.~Park, T.~Prokopec and R.P.~Woodard, \emph{{Quantum Scalar Corrections to the
  Gravitational Potentials on de Sitter Background}},
  \href{https://doi.org/10.1007/JHEP01(2016)074}{\emph{JHEP} {\bfseries 01}
  (2016) 074} [\href{https://arxiv.org/abs/1510.03352}{{\ttfamily
  1510.03352}}].

\bibitem{froebverdaguer2016a}
M.B.~Fr{\"o}b and E.~Verdaguer, \emph{{Quantum corrections to the gravitational
  potentials of a point source due to conformal fields in de Sitter}},
  \href{https://doi.org/10.1088/1475-7516/2016/03/015}{\emph{JCAP} {\bfseries
  1603} (2016) 015} [\href{https://arxiv.org/abs/1601.03561}{{\ttfamily
  1601.03561}}].

\bibitem{froebverdaguer2017}
M.B.~Fr{\"o}b and E.~Verdaguer, \emph{{Quantum corrections for spinning
  particles in de Sitter}},
  \href{https://doi.org/10.1088/1475-7516/2017/04/022}{\emph{JCAP} {\bfseries
  04} (2017) 022} [\href{https://arxiv.org/abs/1701.06576}{{\ttfamily
  1701.06576}}].

\bibitem{mukhanov}
V.~Mukhanov, \emph{{Physical Foundations of Cosmology}},
  \href{http://www.worldcat.org/search?q=isbn:9780521563987}{ Cambridge
  University Press, Cambridge, UK } (2005).

\bibitem{kazanas1980}
D.~Kazanas, \emph{{Dynamics of the Universe and Spontaneous Symmetry
  Breaking}}, \href{https://doi.org/10.1086/183361}{\emph{Astrophys.~J.}
  {\bfseries 241} (1980) L59}.

\bibitem{sato1981}
K.~Sato, \emph{{First order phase transition of a vacuum and expansion of the
  Universe}},
  \href{https://doi.org/https://doi.org/10.1093/mnras/195.3.467}{\emph{Mon.~Not.~Roy.~Astron.~Soc.}
  {\bfseries 195} (1981) 467}.

\bibitem{guth1981}
A.H.~Guth, \emph{{The Inflationary Universe: a possible solution to the horizon
  and flatness problems}},
  \href{https://doi.org/10.1103/PhysRevD.23.347}{\emph{Phys.~Rev.~D} {\bfseries
  23} (1981) 347}.

\bibitem{linde1982}
A.D.~Linde, \emph{{A new inflationary universe scenario: a possible solution of
  the horizon, flatness, homogeneity, isotropy, and primordial monopole
  problems}},
  \href{https://doi.org/10.1016/0370-2693(82)91219-9}{\emph{Phys.~Lett.~B}
  {\bfseries 108} (1982) 389}.

\bibitem{albrechtsteinhardt1982}
A.~Albrecht and P.J.~Steinhardt, \emph{{Cosmology for grand unified theories
  with radiatively induced symmetry breaking}},
  \href{https://doi.org/10.1103/PhysRevLett.48.1220}{\emph{Phys.~Rev.~Lett.}
  {\bfseries 48} (1982) 1220}.

\bibitem{perlmutteretal1998}
{\scshape Supernova Cosmology Project} collaboration, \emph{{Measurements of
  $\Omega$ and $\Lambda$ from 42 high redshift supernovae}},
  \href{https://doi.org/10.1086/307221}{\emph{Astrophys.~J.} {\bfseries 517}
  (1999) 565} [\href{https://arxiv.org/abs/astro-ph/9812133}{{\ttfamily
  astro-ph/9812133}}].

\bibitem{knopetal2003}
{\scshape Supernova Cosmology Project} collaboration, \emph{{New constraints on
  $\Omega_{\mathrm{M}}$, $\Omega_\Lambda$, and $w$ from an independent set of
  eleven high-redshift supernovae observed with HST}},
  \href{https://doi.org/10.1086/378560}{\emph{Astrophys.~J.} {\bfseries 598}
  (2003) 102} [\href{https://arxiv.org/abs/astro-ph/0309368}{{\ttfamily
  astro-ph/0309368}}].

\bibitem{eisensteinetal2005}
{\scshape SDSS} collaboration, \emph{{Detection of the baryon acoustic peak in
  the large-scale correlation function of SDSS luminous red galaxies}},
  \href{https://doi.org/10.1086/466512}{\emph{Astrophys.~J.} {\bfseries 633}
  (2005) 560} [\href{https://arxiv.org/abs/astro-ph/0501171}{{\ttfamily
  astro-ph/0501171}}].

\bibitem{astieretal2006}
{\scshape SNLS} collaboration, \emph{{The Supernova legacy survey: Measurement
  of $\Omega_{\mathrm{M}}$, $\Omega_\Lambda$ and $w$ from the first year data
  set}},
  \href{https://doi.org/10.1051/0004-6361:20054185}{\emph{Astron.~Astrophys.}
  {\bfseries 447} (2006) 31}
  [\href{https://arxiv.org/abs/astro-ph/0510447}{{\ttfamily
  astro-ph/0510447}}].

\bibitem{kowalskietal2008}
{\scshape Supernova Cosmology Project} collaboration, \emph{{Improved
  Cosmological Constraints from New, Old and Combined Supernova Datasets}},
  \href{https://doi.org/10.1086/589937}{\emph{Astrophys.~J.} {\bfseries 686}
  (2008) 749} [\href{https://arxiv.org/abs/0804.4142}{{\ttfamily 0804.4142}}].

\bibitem{bardeen1980}
J.M.~Bardeen, \emph{{Gauge-invariant cosmological perturbations}},
  \href{https://doi.org/10.1103/PhysRevD.22.1882}{\emph{Phys.~Rev.~D}
  {\bfseries 22} (1980) 1882}.

\bibitem{brunettietal2016}
R.~Brunetti, K.~Fredenhagen, T.-P.~Hack, N.~Pinamonti and K.~Rejzner,
  \emph{{Cosmological perturbation theory and quantum gravity}},
  \href{https://doi.org/10.1007/JHEP08(2016)032}{\emph{JHEP} {\bfseries 08}
  (2016) 032} [\href{https://arxiv.org/abs/1605.02573}{{\ttfamily
  1605.02573}}].

\bibitem{froeb2018}
M.B.~Fr{\"o}b, \emph{{Gauge-invariant quantum gravitational corrections to
  correlation functions}},
  \href{https://doi.org/10.1088/1361-6382/aaa74c}{\emph{Class.~Quant.~Grav.}
  {\bfseries 35} (2018) 055006}
  [\href{https://arxiv.org/abs/1710.00839}{{\ttfamily 1710.00839}}].

\bibitem{froeblima2018}
M.B.~Fr{\"o}b and W.C.C.~Lima, \emph{{Propagators for gauge-invariant
  observables in cosmology}},
  \href{https://doi.org/10.1088/1361-6382/aab427}{\emph{Class.~Quant.~Grav.}
  {\bfseries 35} (2018) 095010}
  [\href{https://arxiv.org/abs/1711.08470}{{\ttfamily 1711.08470}}].

\bibitem{froeb2019}
M.B.~Fr{\"o}b, \emph{{One-loop quantum gravitational backreaction on the local
  Hubble rate}},
  \href{https://doi.org/10.1088/1361-6382/ab10fb}{\emph{Class.~Quant.~Grav.}
  {\bfseries 36} (2019) 095010}
  [\href{https://arxiv.org/abs/1806.11124}{{\ttfamily 1806.11124}}].

\bibitem{lima2020}
W.C.C.~Lima, \emph{{Graviton backreaction on the local cosmological expansion
  in slow-roll inflation}},
  \href{https://doi.org/10.1088/1361-6382/abfaeb}{\emph{Class.~Quantum~Grav.}
  {\bfseries 38} (2021) 135015}
  [\href{https://arxiv.org/abs/2007.04995}{{\ttfamily 2007.04995}}].

\bibitem{komar1958}
A.~Komar, \emph{{Construction of a Complete Set of Independent Observables in
  the General Theory of Relativity}},
  \href{https://doi.org/10.1103/PhysRev.111.1182}{\emph{Phys.~Rev.} {\bfseries
  111} (1958) 1182}.

\bibitem{bergmannkomar1960}
P.G.~Bergmann and A.B.~Komar, \emph{{Poisson brackets between locally defined
  observables in general relativity}},
  \href{https://doi.org/10.1103/PhysRevLett.4.432}{\emph{Phys.~Rev.~Lett.}
  {\bfseries 4} (1960) 432}.

\bibitem{bergmann1961}
P.G.~Bergmann, \emph{{Observables in General Relativity}},
  \href{https://doi.org/10.1103/RevModPhys.33.510}{\emph{Rev.~Mod.~Phys.}
  {\bfseries 33} (1961) 510}.

\bibitem{gieseletal2010a}
K.~Giesel, S.~Hofmann, T.~Thiemann and O.~Winkler, \emph{{Manifestly
  Gauge-Invariant General Relativistic Perturbation Theory. I. Foundations}},
  \href{https://doi.org/10.1088/0264-9381/27/5/055005}{\emph{Class.~Quantum~Grav.}
  {\bfseries 27} (2010) 055005}
  [\href{https://arxiv.org/abs/0711.0115}{{\ttfamily 0711.0115}}].

\bibitem{gieseletal2010b}
K.~Giesel, S.~Hofmann, T.~Thiemann and O.~Winkler, \emph{{Manifestly
  Gauge-invariant general relativistic perturbation theory. II. FRW background
  and first order}},
  \href{https://doi.org/10.1088/0264-9381/27/5/055006}{\emph{Class.~Quantum~Grav.}
  {\bfseries 27} (2010) 055006}
  [\href{https://arxiv.org/abs/0711.0117}{{\ttfamily 0711.0117}}].

\bibitem{tambornino2012}
J.~Tambornino, \emph{{Relational Observables in Gravity: a Review}},
  \href{https://doi.org/10.3842/SIGMA.2012.017}{\emph{SIGMA} {\bfseries 8}
  (2012) 017} [\href{https://arxiv.org/abs/1109.0740}{{\ttfamily 1109.0740}}].

\bibitem{brownkuchar1995}
J.D.~Brown and K.V.~Kucha{\v r}, \emph{{Dust as a standard of space and time in
  canonical quantum gravity}},
  \href{https://doi.org/10.1103/PhysRevD.51.5600}{\emph{Phys.~Rev.~D}
  {\bfseries 51} (1995) 5600}
  [\href{https://arxiv.org/abs/gr-qc/9409001}{{\ttfamily gr-qc/9409001}}].

\bibitem{mtw}
C.~Misner, K.~Thorne and J.A.~Wheeler, \emph{{Gravitation}},
  \href{http://www.worldcat.org/search?q=isbn:9780716703440}{ W.~H.~Freeman,
  San Francisco } (1973).

\bibitem{stewartwalker1974}
J.M.~Stewart and M.~Walker, \emph{{Perturbations of space-times in general
  relativity}},
  \href{https://doi.org/10.1098/rspa.1974.0172}{\emph{Proc.~R.~Soc.~A}
  {\bfseries 341} (1974) 49}.

\bibitem{canepadappiaggikhavkine2018}
G.~Canepa, C.~Dappiaggi and I.~Khavkine, \emph{{IDEAL characterization of
  isometry classes of FLRW and inflationary spacetimes}},
  \href{https://doi.org/10.1088/1361-6382/aa9f61}{\emph{Class.~Quant.~Grav.}
  {\bfseries 35} (2018) 035013}
  [\href{https://arxiv.org/abs/1704.05542}{{\ttfamily 1704.05542}}].

\bibitem{froebhackkhavkine2018}
M.B.~Fr{\"o}b, T.-P.~Hack and I.~Khavkine, \emph{{Approaches to linear local
  gauge-invariant observables in inflationary cosmologies}},
  \href{https://doi.org/10.1088/1361-6382/aabcb7}{\emph{Class.~Quant.~Grav.}
  {\bfseries 35} (2018) 115002}
  [\href{https://arxiv.org/abs/1801.02632}{{\ttfamily 1801.02632}}].

\bibitem{khavkine2019}
I.~Khavkine, \emph{{IDEAL characterization of higher dimensional spherically
  symmetric black holes}},
  \href{https://doi.org/10.1088/1361-6382/aafcf1}{\emph{Class.~Quant.~Grav.}
  {\bfseries 36} (2019) 045001}
  [\href{https://arxiv.org/abs/1807.09699}{{\ttfamily 1807.09699}}].

\bibitem{leibbrandt1975}
G.~Leibbrandt, \emph{{Introduction to the Technique of Dimensional
  Regularization}},
  \href{https://doi.org/10.1103/RevModPhys.47.849}{\emph{Rev.~Mod.~Phys.}
  {\bfseries 47} (1975) 849}.

\bibitem{schroedinger1930}
E.~Schr{\"o}dinger, \emph{{\"Uber die kr\"aftefreie Bewegung in der
  relativistischen Quantenmechanik}}, {\emph{Sitzungsber. Preu{\ss}. Akad.
  Wiss., Phys.-Math. Klasse} {\bfseries 24} (1930) 418}.

\bibitem{thaller}
B.~Thaller, \emph{{The Dirac Equation}},
  \href{http://www.worldcat.org/search?q=isbn:3540548331}{ Springer-Verlag,
  Berlin, Heidelberg, New York } (1992).

\bibitem{kobakhidzemanningtureanu2016}
A.~Kobakhidze, A.~Manning and A.~Tureanu, \emph{{Observable Zitterbewegung in
  Curved Spacetimes}},
  \href{https://doi.org/10.1016/j.physletb.2016.03.049}{\emph{Phys.~Lett.~B}
  {\bfseries 757} (2016) 84}
  [\href{https://arxiv.org/abs/1508.06322}{{\ttfamily 1508.06322}}].

\bibitem{ecksteinfrancomiller2017}
M.~Eckstein, N.~Franco and T.~Miller, \emph{{Noncommutative geometry of
  Zitterbewegung}},
  \href{https://doi.org/10.1103/PhysRevD.95.061701}{\emph{Phys.~Rev.~D}
  {\bfseries 95} (2017) 061701}
  [\href{https://arxiv.org/abs/1610.10083}{{\ttfamily 1610.10083}}].

\bibitem{becchirouetstora1975}
C.~Becchi, A.~Rouet and R.~Stora, \emph{{Renormalization of the Abelian
  Higgs-Kibble Model}},
  \href{https://doi.org/10.1007/BF01614158}{\emph{Commun.~Math.~Phys.}
  {\bfseries 42} (1975) 127}.

\bibitem{becchirouetstora1976}
C.~Becchi, A.~Rouet and R.~Stora, \emph{{Renormalization of Gauge Theories}},
  \href{https://doi.org/10.1016/0003-4916(76)90156-1}{\emph{Ann.~Phys.}
  {\bfseries 98} (1976) 287}.

\bibitem{kugoojima1978}
T.~Kugo and I.~Ojima, \emph{{Manifestly Covariant Canonical Formulation of
  Yang-Mills Field Theories. I. General Formalism}},
  \href{https://doi.org/10.1143/PTP.60.1869}{\emph{Prog.~Theor.~Phys.}
  {\bfseries 60} (1978) 1869}.

\bibitem{weinberg_v2}
S.~Weinberg, \emph{{The Quantum Theory of Fields, Volume 2: Modern
  Applications}}, \href{http://www.worldcat.org/search?q=isbn:9780521670548}{
  Cambridge University Press, Cambridge, UK } (2005).

\bibitem{barnichbrandthenneaux2000}
G.~Barnich, F.~Brandt and M.~Henneaux, \emph{{Local BRST cohomology in gauge
  theories}},
  \href{https://doi.org/10.1016/S0370-1573(00)00049-1}{\emph{Phys.~Rept.}
  {\bfseries 338} (2000) 439}
  [\href{https://arxiv.org/abs/hep-th/0002245}{{\ttfamily hep-th/0002245}}].

\bibitem{cappernamazie1978}
D.M.~Capper and M.A.~Namazie, \emph{{A general gauge calculation of the
  graviton self-energy}},
  \href{https://doi.org/10.1016/0550-3213(78)90229-8}{\emph{Nucl.~Phys.~B}
  {\bfseries 142} (1978) 535}.

\bibitem{capper1980}
D.M.~Capper, \emph{{A general gauge graviton loop calculation}},
  \href{https://doi.org/10.1088/0305-4470/13/1/022}{\emph{J.~Phys.~A}
  {\bfseries 13} (1980) 199}.

\bibitem{hsukim2012}
J.-P.~Hsu and S.H.~Kim, \emph{{The S-matrix and graviton self-energy in quantum
  Yang-Mills gravity}},
  \href{https://doi.org/10.1140/epjp/i2012-12146-3}{\emph{Eur.~Phys.~J.~Plus}
  {\bfseries 127} (2012) 146}
  [\href{https://arxiv.org/abs/1210.4503}{{\ttfamily 1210.4503}}].

\bibitem{xact}
J.M.~Mart{\'\i}n-Garc{\'\i}a, A.~Garc{\'\i}a-Parrado, A.~Stecchina, B.~Wardell,
  C.~Pitrou, D.~Brizuela et~al., ``{xAct: Efficient tensor computer algebra for
  the Wolfram Language}.'' \href{http://www.xact.es}{http://www.xact.es}, 2020.

\bibitem{smirnov2004}
V.A.~Smirnov, \emph{{Evaluating Feynman integrals}}, vol.~211 of
  \emph{{Springer Tracts in Modern Physics}},
  \href{http://www.worldcat.org/search?q=isbn:9783540239338}{
  {Springer-Verlag}, {Berlin/Heidelberg, Germany} } (2004),
  \href{https://doi.org/10.1007/b95498}{10.1007/b95498}.

\bibitem{blanchetdamourespositofarese2004}
L.~Blanchet, T.~Damour and G.~Esposito-Far{\`e}se, \emph{{Dimensional
  regularization of the third post-Newtonian dynamics of point particles in
  harmonic coordinates}},
  \href{https://doi.org/10.1103/PhysRevD.69.124007}{\emph{Phys.~Rev.~D}
  {\bfseries 69} (2004) 124007}
  [\href{https://arxiv.org/abs/gr-qc/0311052}{{\ttfamily gr-qc/0311052}}].

\bibitem{blanchetetal2005}
L.~Blanchet, T.~Damour, G.~Esposito-Far{\`e}se and B.R.~Iyer,
  \emph{{Dimensional regularization of the third post-Newtonian gravitational
  wave generation from two point masses}},
  \href{https://doi.org/10.1103/PhysRevD.71.124004}{\emph{Phys.~Rev.~D}
  {\bfseries 71} (2005) 124004}
  [\href{https://arxiv.org/abs/gr-qc/0503044}{{\ttfamily gr-qc/0503044}}].

\bibitem{jaranowskischaefer2013}
P.~Jaranowski and G.~Sch{\"a}fer, \emph{{Dimensional regularization of local
  singularities in the 4th post-Newtonian two-point-mass Hamiltonian}},
  \href{https://doi.org/10.1103/PhysRevD.87.081503}{\emph{Phys.~Rev.~D}
  {\bfseries 87} (2013) 081503}
  [\href{https://arxiv.org/abs/1303.3225}{{\ttfamily 1303.3225}}].

\bibitem{bernardetal2015}
L.~Bernard, L.~Blanchet, A.~Boh{\'e}, G.~Faye and S.~Marsat, \emph{{Fokker
  action of nonspinning compact binaries at the fourth post-Newtonian
  approximation}},
  \href{https://doi.org/10.1103/PhysRevD.93.084037}{\emph{Phys.~Rev.~D}
  {\bfseries 93} (2016) 084037}
  [\href{https://arxiv.org/abs/1512.02876}{{\ttfamily 1512.02876}}].

\bibitem{duff1973}
M.J.~Duff, \emph{{Quantum Tree Graphs and the Schwarzschild Solution}},
  \href{https://doi.org/10.1103/PhysRevD.7.2317}{\emph{Phys.~Rev.~D} {\bfseries
  7} (1973) 2317}.

\bibitem{levi2018}
M.~Levi, \emph{{Effective Field Theories of Post-Newtonian Gravity: A
  comprehensive review}},
  \href{https://doi.org/10.1088/1361-6633/ab12bc}{\emph{Rept. Prog. Phys.}
  {\bfseries 83} (2020) 075901}
  [\href{https://arxiv.org/abs/1807.01699}{{\ttfamily 1807.01699}}].

\bibitem{tomboulis1977}
E.~Tomboulis, \emph{{$1/N$ expansion and renormalization in quantum gravity}},
  \href{https://doi.org/10.1016/0370-2693(77)90678-5}{\emph{Phys.~Lett.~B}
  {\bfseries 70} (1977) 361}.

\bibitem{hartlehorowitz1981}
J.B.~Hartle and G.T.~Horowitz, \emph{{Ground-state expectation value of the
  metric in the $1/N$ or semiclassical approximation to quantum gravity}},
  \href{https://doi.org/10.1103/PhysRevD.24.257}{\emph{Phys.~Rev.~D} {\bfseries
  24} (1981) 257}.

\bibitem{hurouraverdaguer2004}
B.L.~Hu, A.~Roura and E.~Verdaguer, \emph{{Induced quantum metric fluctuations
  and the validity of semiclassical gravity}},
  \href{https://doi.org/10.1103/PhysRevD.70.044002}{\emph{Phys.~Rev.~D}
  {\bfseries 70} (2004) 044002}
  [\href{https://arxiv.org/abs/gr-qc/0402029}{{\ttfamily gr-qc/0402029}}].

\bibitem{gasperinietal2011}
M.~Gasperini, G.~Marozzi, F.~Nugier and G.~Veneziano, \emph{{Light-cone
  averaging in cosmology: Formalism and applications}},
  \href{https://doi.org/10.1088/1475-7516/2011/07/008}{\emph{JCAP} {\bfseries
  07} (2011) 008} [\href{https://arxiv.org/abs/1104.1167}{{\ttfamily
  1104.1167}}].

\bibitem{froeblima2021}
M.B.~Fr{\"o}b and W.C.C.~Lima, \emph{{Cosmological Perturbations and Invariant
  Observables in Geodesic Lightcone Coordinates}},
  \href{https://arxiv.org/abs/2108.11960}{{\ttfamily 2108.11960}}.

\bibitem{modesto2005}
L.~Modesto, \emph{{Loop quantum black hole}},
  \href{https://doi.org/10.1088/0264-9381/23/18/006}{\emph{Class.~Quant.~Grav.}
  {\bfseries 23} (2006) 5587}
  [\href{https://arxiv.org/abs/gr-qc/0509078}{{\ttfamily gr-qc/0509078}}].

\bibitem{tsamiswoodard1996}
N.C.~Tsamis and R.P.~Woodard, \emph{{Quantum gravity slows inflation}},
  \href{https://doi.org/10.1016/0550-3213(96)00246-5}{\emph{Nucl.~Phys.~B}
  {\bfseries 474} (1996) 235}
  [\href{https://arxiv.org/abs/hep-ph/9602315}{{\ttfamily hep-ph/9602315}}].

\end{thebibliography}\endgroup
  
\end{document}